\documentclass[a4paper,11pt]{article}

\usepackage[a4paper,margin=2.7cm]{geometry}
\usepackage[T1]{fontenc}
\usepackage{lmodern}
\usepackage{microtype}
\usepackage{amsmath,amssymb,amsfonts,amsthm}
\usepackage{mathtools}
\usepackage{booktabs}
\usepackage{array}
\usepackage{authblk}
\usepackage{cite}
\usepackage{graphicx}
\usepackage{hyperref}
\usepackage{bookmark}
\usepackage{enumitem}
\usepackage{float}
\usepackage{caption}
\hypersetup{
  colorlinks=true,
  linkcolor=blue,
  citecolor=blue,
  urlcolor=blue,
  pdftitle={One-loop quantum mass and Heun spectrum of the Christ--Lee kink},
  pdfauthor={Mulyanto},
  bookmarksnumbered=true
}

\numberwithin{equation}{section}
\allowdisplaybreaks[1]
\setlist{nosep}

\newcommand{\sech}{\operatorname{sech}}

\newcommand{\HeunG}{\operatorname{HeunG}}

\newcommand{\dd}{\mathrm{d}}
\newcommand{\eps}{\epsilon}
\newcommand{\arsinh}{\operatorname{arsinh}}
\newcommand{\Tr}{\operatorname{Tr}}
\newcommand{\mc}{M_{\mathrm{cl}}}
\newcommand{\arxiv}[1]{\href{https://arxiv.org/abs/#1}{arXiv:#1}}
\newcommand{\doi}[1]{\href{https://doi.org/#1}{doi:#1}}

\begin{document}

\title{\bfseries Heun Spectrum, Threshold Hierarchy, and One-Loop Quantum Mass of the Christ--Lee Kink}
\author[1]{Mulyanto\thanks{Corresponding author: muly031@brin.go.id}}
\author[2,3]{Afif J. Pradipta}
\author[1]{Ardian N. Atmaja}
\author[2]{Bobby E. Gunara}

\affil[1]{\small Research Center for Quantum Physics, National Research and Innovation Agency (BRIN), Kompleks PUSPIPTEK Serpong, Tangerang 15310, Indonesia}
\affil[2]{\small Theoretical Physics Laboratory, Theoretical High Energy Physics Research Division, Faculty of Mathematics and Natural Sciences, Institut Teknologi Bandung, Jl. Ganesha no. 10, Bandung 40132, Indonesia}
\affil[3]{\small School of Physical Sciences, University of Science and Technology of China, Hefei 230026, China}

\date{}
\maketitle

\begin{abstract}
We study the fluctuation spectrum and renormalized one-loop quantum
mass of the Christ--Lee kink over the full positive range of its
deformation parameter. The fluctuation equation is reduced exactly to
general-Heun form, yielding parity-resolved continuum solutions and
implicit quantization conditions for the discrete spectrum. At the
first nontrivial continuum crossing, the odd Heun solution truncates to
a polynomial, giving the exact threshold
\(\epsilon_1=\sqrt{\sqrt{3}-1}\). For small deformation, the kink separates into two widely spaced
\(\phi^6\)-like interfaces and the fluctuation operator develops a long
central cavity. This geometry explains the logarithmic growth of the
bound-state count, the asymptotically geometric hierarchy of
continuum-threshold crossings, and the emergence of a soft
relative-translation mode. In the opposite limit, the spectrum
approaches the \(\phi^4\) P\"oschl--Teller problem, while its even
threshold resonance becomes a shallow bound state at finite
deformation. The one-loop correction is evaluated in a fixed
vacuum-normal-ordering prescription using a common-regulator spectral
trace and independently through an imaginary-frequency functional
determinant. The two calculations agree throughout representative
finite deformations and reproduce the exact \(\phi^4\) limit. In the
small-deformation regime, the extended central region produces a
logarithmically enhanced negative quantum correction. These results
provide a unified analytic and numerical description of how the
spectral reorganization of the Christ--Lee kink controls its
semiclassical quantum mass.
\end{abstract}

\medskip

\tableofcontents

% ============================================================
% ============================================================
\section{Introduction}
\label{sec:introduction}

Topological solitons provide a particularly useful setting in which
nonperturbative classical configurations can be studied together with
systematic quantum corrections. For a static kink, the classical mass
is determined by a nonlinear boundary-value problem, while the leading
quantum correction is governed by the spectrum of the linearized
fluctuation operator. This connection lies at the heart of
semiclassical soliton quantization and continues to provide a useful
laboratory for nonperturbative quantum field theory in low dimensions
\cite{manton,dhn,christ,gervais,graham2022,evslin2026}. In practice,
however, determining the fluctuation spectrum is only part of the
problem. The kink and vacuum sectors must be compared within the same
ultraviolet prescription, and the translational zero mode has to be
treated consistently throughout the calculation.

The standard $\phi^4$ kink provides the simplest benchmark because its
fluctuation operator is the reflectionless P\"oschl--Teller
Hamiltonian~\cite{pt}. Its bound states and scattering data are known
explicitly, and the one-loop mass can be obtained in several equivalent
ways, including direct mode sums~\cite{dhn,ccg}, phase-shift methods
with Born subtraction~\cite{graham1998,sumrules}, and heat-kernel or
zeta-function techniques~\cite{bordag,vassilevich,alonso2011,alonso2012}.
These approaches also emphasize that the treatment of the zero mode,
the ultraviolet subtraction and the renormalization prescription are
not independent choices~\cite{goldhaber,alonsozero}. More recent
Hamiltonian formulations based on vacuum normal ordering make these
cancellations particularly transparent and provide a convenient
framework for quantum states constructed around soliton backgrounds
\cite{evslin2019,normal,zerostates,twoloop,phi4two}.

The Christ--Lee model~\cite{christ} provides a considerably richer
deformation of this benchmark. It is a sextic scalar theory whose kink
interpolates continuously between a $\phi^4$-like configuration at
large deformation and a pair of increasingly separated $\phi^6$-like
interfaces when the deformation parameter becomes small. Related
polynomial and deformed defects have been studied extensively
\cite{lohe,bazeia}. Their internal spectra are relevant not only for
semiclassical quantization but also for nonlinear dynamics, since
localized and near-threshold modes can participate in resonant energy
exchange during kink interactions. This mechanism is familiar from
the $\phi^4$ and $\phi^6$ models~\cite{campbell,dorey2011}. For the
Christ--Lee family in particular, the classical mode spectrum and the
dynamics of weakly bound subkinks were investigated in
Ref.~\cite{dorey2023}. More recently, threshold, antibound and
quasinormal modes in an extended Christ--Lee parametrization have also
been used to organize nonlinear oscillon excitations
\cite{blaschke2026}.

The static spectral problem becomes especially interesting when the
two interfaces separate. The region between them grows while the
fluctuation operator develops a scale different from that of the
asymptotic vacuum. This separation leads to a systematic
reorganization of the discrete spectrum, the emergence of new bound
states through continuum-threshold crossings, and a distinct
near-threshold structure in the opposite $\phi^4$-like limit. The same
deformation-dependent spectral flow also determines how the
renormalized one-loop kink energy evolves across the parameter space. The one-loop mass of this sextic family has previously been
studied using heat-kernel techniques \cite{alonso2011}. Related
heat-kernel methods have also been applied to other solvable
kink families \cite{alonso2012}, providing useful methodological
background for semiclassical mass calculations. The classical
spectral problem has also been explored numerically \cite{dorey2023}.

The purpose of the present work is therefore not to
reintroduce the model or merely to provide another numerical spectrum.
Instead, we develop a unified spectral formulation that connects the
analytic structure of the fluctuation equation, the appearance of
threshold states, the two-interface geometry and the renormalized
quantum mass.

Our strategy combines exact analytic reduction, direct spectral
numerics and asymptotic analysis. We first transform the linear
fluctuation equation to the general Heun equation and construct
solutions of definite parity. Bound-state normalizability is thereby
formulated as a connection problem between the regular solution at the
center of the kink and the decaying Jost branch at spatial infinity.
For the numerical spectrum we do not attempt to evaluate global Heun
connection coefficients directly. Instead, the discrete eigenvalues
and continuum-threshold crossings are determined independently by
shooting. The Heun representation nevertheless retains direct analytic
content: at the first nontrivial threshold the relevant Heun series
terminates, producing an exact polynomial threshold state and providing
an analytic benchmark for the numerical spectral calculation.

We then analyze the two limiting deformation regimes using methods
adapted to their different geometries. For small $\epsilon$, the kink
separates into two widely spaced $\phi^6$-like interfaces and the
fluctuation potential develops an extended central plateau. We treat
this region as an effective one-dimensional cavity. A standing-wave
quantization condition then explains the increasing number of bound
states and the hierarchy of threshold crossings, while the overlap of
the two interface translation modes accounts for an additional soft
relative mode. In this way, the discrete spectrum can be understood
directly from the emerging two-interface geometry rather than only
from a numerical eigenvalue table. The opposite limit is organized around the exactly solvable $\phi^4$
P\"oschl--Teller problem. Expanding the fluctuation operator at large
$\epsilon$ allows the near-threshold spectrum to be studied
perturbatively. In particular, the even threshold resonance of the
limiting $\phi^4$ problem is followed as it moves below the continuum
at finite deformation. The two asymptotic regimes therefore involve
different spectral mechanisms: confinement by a long intermediate
region at small deformation and perturbation of a threshold resonance
at large deformation.

Finally, we compute the one-loop kink energy in a fixed
vacuum-normal-ordering, or no-tadpole, prescription. The kink and
vacuum sectors are regulated together so that bound states, continuum
modes and the translational zero mode enter a single finite spectral
trace. We evaluate this representation numerically using spatial
discretization followed by continuum extrapolation. As an independent
check, we construct an imaginary-frequency functional determinant
directly from the fluctuation equation, with the corresponding Born
subtraction chosen to implement the same renormalization prescription.
Agreement between these two calculations, together with recovery of
the exactly known $\phi^4$ limit, provides a nontrivial validation of
the finite-deformation quantum correction. The same framework also
allows the small- and large-deformation behavior of the one-loop mass
to be related directly to the corresponding spectral geometry.

The paper is organized as follows. Section~\ref{sec:model} introduces
the Christ--Lee model, its kink solution and the corresponding
semiclassical fluctuation problem. Section~\ref{sec:heun} develops the
general-Heun representation of the parity modes. In
Sec.~\ref{sec:bound-states} we determine the discrete spectrum and
threshold crossings and analyze the small- and large-deformation
limits. Section~\ref{sec:one-loop} constructs the renormalized one-loop
energy, compares the spectral-trace and functional-determinant
representations, and studies its deformation dependence.
Section~\ref{sec:conclusion} summarizes the main results. Technical
derivations and numerical details are collected in the appendices.

\section{Christ--Lee model and semiclassical kink sector}
\label{sec:model}

\subsection{Classical model, vacuum structure and kink}
\label{subsec:vacuum}

We begin with a real scalar field $\Phi(t,x)$ in $1+1$ dimensions,
described by the Lagrangian density
\begin{equation}
\mathcal{L}
=
\frac{1}{2}(\partial_t\Phi)^2
-
\frac{1}{2}(\partial_x\Phi)^2
-
V(\Phi),
\label{eq:lagrangian}
\end{equation}
where
\begin{equation}
V(\Phi)
=
\frac{\lambda}{4v^2(1+\epsilon^2)}
\left(\epsilon^2v^2+\Phi^2\right)
\left(v^2-\Phi^2\right)^2,
\qquad
\lambda>0,
\quad
v>0,
\quad
\epsilon\geq0.
\label{eq:potential}
\end{equation}
The parameter $v$ fixes the field scale and $\lambda$ sets the overall mass scale. The dimensionless parameter $\epsilon$ controls the
deformation of the model and, as we shall see below, also governs the
separation of the two interfaces that emerge in the small-$\epsilon$
regime. Varying Eq.~\eqref{eq:lagrangian} with respect to $\Phi$ gives
\begin{equation}
\partial_t^2\Phi
-
\partial_x^2\Phi
+
\frac{\lambda}{2v^2(1+\epsilon^2)}
\Phi\left(v^2-\Phi^2\right)
\left(v^2-2\epsilon^2v^2-3\Phi^2\right)
=0.
\label{eq:classical-eom}
\end{equation}
Before discussing the kink itself, it is useful to look at the vacuum
structure of the theory.  For every finite $\epsilon>0$, the two
degenerate true vacua are
\begin{equation}
\Phi_{\pm}=\pm v,
\qquad
V(\pm v)=0.
\label{eq:true-vacua}
\end{equation}
Expanding the potential about either of these vacua gives the mass of
the elementary fluctuation,
\begin{equation}
m^2\equiv V''(\pm v)=2\lambda v^2,
\qquad
m=\sqrt{2\lambda}\,v.
\label{eq:vacuum-mass}
\end{equation}
The point $\Phi=0$ plays a different role.  At finite $\epsilon$ one
finds
\begin{equation}
V(0)=\frac{\lambda\epsilon^2v^4}{4(1+\epsilon^2)},
\qquad
V''(0)=\frac{m^2}{4}\frac{1-2\epsilon^2}{1+\epsilon^2}.
\label{eq:central-vacuum-data}
\end{equation}
Thus, for $0<\epsilon<1/\sqrt2$, $\Phi=0$ is a local secondary (false) vacuum with positive energy density.  This point becomes exactly degenerate with $\Phi=\pm v$ only in the strict $\epsilon=0$ theory.  These facts will later explain why the small-$\epsilon$ kink develops a long central region with a different fluctuation mass.

For $\epsilon>0$, the model admits a static kink interpolating directly between the two true vacua,
\begin{equation}
\Phi_K(-\infty)=-v,
\qquad
\Phi_K(+\infty)=+v,
\end{equation}
with profile
\begin{equation}
\Phi_K(x)
=
\frac{v\epsilon\sinh(mx/2)}
{\sqrt{1+\epsilon^2\cosh^2(mx/2)}}.
\label{eq:unshifted-kink}
\end{equation}
It is worth separating this solution from the strict $\epsilon=0$
theory.  At $\epsilon=0$, the points $-v$, $0$, and $+v$ are three
degenerate vacua, so the elementary topological sectors are
$(-v,0)$ and $(0,+v)$ rather than $(-v,+v)$.  A convenient representative of the $(0,+v)$ sector is
\begin{equation}
\Phi_{0+}(x)
=
v\sqrt{\frac{1+\tanh(mx/2)}{2}}
=
\frac{v}{\sqrt{1+e^{-mx}}}.
\label{eq:epsilon-zero-kink}
\end{equation}
The strict $\epsilon=0$ theory must therefore be distinguished from the limiting configuration $\epsilon\to0^+$ of the finite-$\epsilon$ kink in Eq.~\eqref{eq:unshifted-kink}.

Using the static first integral, the classical mass of the finite-$\epsilon$ kink is
\begin{align}
\mc
&=
\int_{-\infty}^{\infty}(\Phi_K')^2\,\dd x
=
\frac{m^3}{2\lambda\sqrt{1+\eps^2}}
\int_0^1(1-s^2)\sqrt{s^2+\eps^2}\,\dd s
\nonumber\\
&=
\frac{m^3}{16\lambda}
\left[
2-\eps^2
+
\frac{\eps^2(\eps^2+4)}{\sqrt{1+\eps^2}}
\arsinh\frac1\eps
\right].
\label{eq:classical}
\end{align}
Here $s=\Phi/v$ is a dimensionless integration variable on the positive half of the kink.  In particular,
$\mc\to m^3/(3\lambda)$ as $\eps\to\infty$ and
$\mc\to m^3/(8\lambda)$ as $\eps\to0^+$.  

\subsubsection{Two-subkink geometry for \texorpdfstring{$\epsilon\to0^+$}{epsilon -> 0+}}
\label{subsubsec:two-subkink}
To make the small-$\epsilon$ geometry transparent, we introduce the dimensionless coordinate,
\begin{equation}
X\equiv mx,
\qquad
\chi_\epsilon(X)\equiv\frac{\Phi_K(x)}{v}.
\label{eq:dimensionless-classical-variables}
\end{equation}
Equation~\eqref{eq:unshifted-kink} then becomes
\begin{equation}
\chi_\epsilon(X)
=
\frac{\eps\sinh(X/2)}{\sqrt{1+\eps^2\cosh^2(X/2)}}.
\label{eq:scaled-kink-small}
\end{equation}
In the same coordinate the elementary $(0,+v)$ kink in Eq.~\eqref{eq:epsilon-zero-kink} is
\begin{equation}
\chi_{0+}(X)=\frac{1}{\sqrt{1+e^{-X}}}.
\label{eq:phi6-elementary}
\end{equation}
We can now see explicitly how the finite-$\epsilon$ kink splits as
$\epsilon$ becomes small. The two transition regions are centered at
\begin{equation}
X_{\pm}=\pm X_0,
\qquad
X_0=2\log\frac{2}{\eps}+o(1),
\label{eq:subkink-centres}
\end{equation}
Near the two transition regions, the asymptotic decomposition is most naturally described in coordinates moving with the interfaces. We define
\begin{equation}
 Y_{+}=X-X_{0},
\qquad
Y_{-}=X+X_{0}.   
\end{equation}
Then, as $\epsilon\to0^{+}$,
\begin{equation}
 \chi_{\epsilon}(X_{0}+Y_{+})
=
\chi_{0+}(Y_{+})+o(1),   
\end{equation}
and
\begin{equation}
 \chi_{\epsilon}(-X_{0}+Y_{-})
=
-\chi_{0+}(-Y_{-})+o(1),   
\end{equation}
where the remainders tend to zero uniformly for $Y_{\pm}$ in compact
sets. Equivalently, in neighborhoods moving with the two interfaces,
\begin{equation}
  \chi_{\epsilon}(X)
=
\chi_{0+}(X-X_{0})
-
\chi_{0+}(-X-X_{0})
+
o(1),
\qquad
\epsilon\to0^{+}.  
\end{equation}
This matched-interface asymptotics should not be interpreted as a
nontrivial pointwise limit at fixed $X$. Since $X_{0}\to\infty$ as
$\epsilon\to0^{+}$, one instead has
$\chi_{\epsilon}(X)\to0$ for every fixed finite $X$.

The first term describes the right interface centered at $+X_0$,
whereas the reflected second term describes the left interface
centered at $-X_0$. It is then natural to define their dimensionless separation by
\begin{equation}
\ell_X \equiv X_+-X_-
       =2X_0
       =4\log\frac{2}{\epsilon}+o(1),
\qquad \epsilon\to0^+.
\label{eq:subkink-separation}
\end{equation}
Thus $\ell_X$ diverges logarithmically as $\epsilon$ approaches zero.

\begin{figure}[H]
    \centering
    \includegraphics[width=0.82\linewidth]{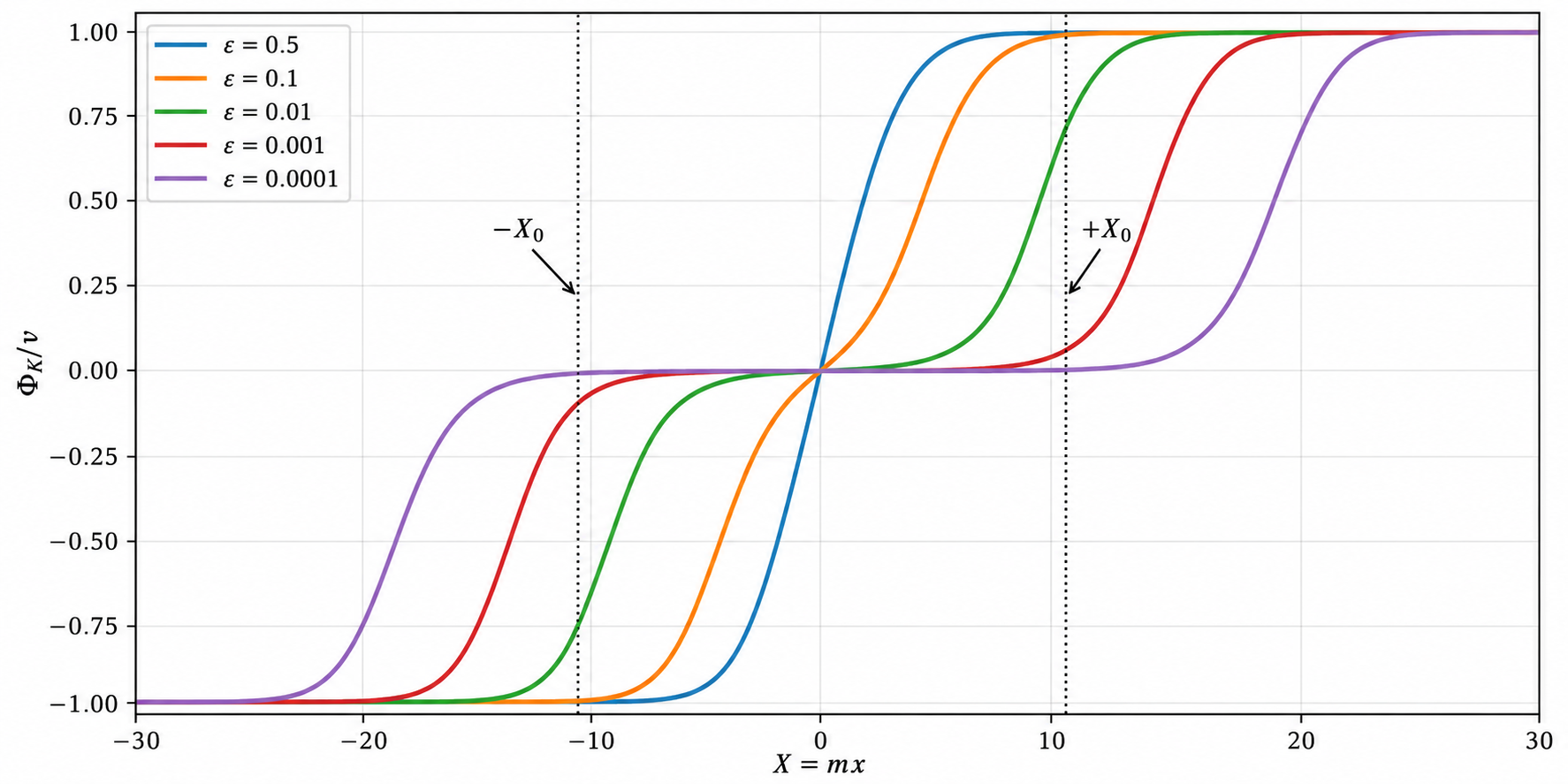}
    \caption{Christ--Lee kink profiles $\Phi_K/v$ in the dimensionless
    coordinate $X=mx$ for several values of the deformation parameter
    $\epsilon$.  As $\epsilon$ decreases, the kink develops two increasingly
    well-separated transition regions with an extended central plateau near
    $\Phi=0$.  The dotted vertical lines mark the leading asymptotic
    subkink-center positions $X=\pm X_0$ for the representative case
    $\epsilon=10^{-2}$, where
    $X_0\simeq2\log(2/\epsilon)\approx10.60$.}
    \label{fig:small-eps-profiles}
\end{figure}

There is a small subtlety in taking this limit.  If $X$ is kept fixed
while $\epsilon\to0^+$, Eq.~\eqref{eq:scaled-kink-small} tends to zero,
because both transition layers are moving out toward
$X=\pm\infty$.  The topological transition has not disappeared; rather,
the limit is nonuniform in $X$.  This distinction is useful later when
we discuss the soft relative mode, the cavity-like part of the
spectrum, and the small-$\epsilon$ behavior of the one-loop energy. Figure~\ref{fig:small-eps-profiles} gives a direct picture of this
separation.  As $\epsilon$ is decreased, the two transition regions move
apart and the field stays close to the intermediate value $\Phi=0$
over an increasingly long interval.  For instance, taking the representative value $\epsilon=10^{-2}$, the leading-order estimate yields
\begin{equation}
X_0 \simeq 2\log\frac{2}{\epsilon}.
\end{equation}

\subsection{Quantization about the \texorpdfstring{$\Phi=-v$}{Phi = -v} vacuum}
\label{subsec:vacuum-quantization}
We next set up the perturbative vacuum sector that will be used to
define the one-loop subtraction.  We choose the left true vacuum,
$\Phi=-v$, and denote the corresponding perturbative state by
$|-\rangle$.  It is convenient to shift the field so that this vacuum
lies at the origin,
\begin{equation}
\phi(x)\equiv\Phi(x)+v,
\qquad
\pi(x)\equiv\partial_t\phi(x)=\partial_t\Phi(x).
\label{eq:vacuum-shift}
\end{equation}
With this convention, the shifted potential is
\begin{equation}
\mathcal V(\phi)\equiv V(\phi-v),
\end{equation}
and the chosen vacuum satisfies $\langle -|\phi(x)|-\rangle=0$.
We then split the Hamiltonian into its free and interacting parts,
\begin{equation}
H=H_0+H_1,
\qquad
H_j=\int_{-\infty}^{\infty}\dd x\,\mathcal H_j(x),
\label{eq:H-decomposition}
\end{equation}
and define the perturbative vacuum energy through
\begin{equation}
H|-\rangle=E_0|-\rangle.
\label{eq:vacuum-eigenvalue}
\end{equation}
Normal ordering will be taken with respect to the vacuum plane-wave
oscillators introduced below.  With the mass $m$ fixed by
Eq.~\eqref{eq:vacuum-mass}, the free Hamiltonian density is
\begin{equation}
\mathcal H_0
=
\frac12:\pi^2:
+
\frac12:(\partial_x\phi)^2:
+
\frac{m^2}{2}:\phi^2:,
\label{eq:H0-density}
\end{equation}
while the remaining terms may be written as
\begin{align}
\mathcal H_1
={}&
C_3:\phi^3:
+C_4:\phi^4:
+C_5:\phi^5:
+C_6:\phi^6:,
\label{eq:H1-density}
\\
C_3
={}&
-m\sqrt{\frac{\lambda}{2}}
\frac{3+\epsilon^2}{1+\epsilon^2},
\qquad
C_4
=
\frac{\lambda}{4}
\frac{13+\epsilon^2}{1+\epsilon^2},
\label{eq:C3-C4}
\\
C_5
={}&
-\frac{3\lambda^{3/2}}
{\sqrt2\,m(1+\epsilon^2)},
\qquad
C_6
=
\frac{\lambda^2}
{2m^2(1+\epsilon^2)}.
\label{eq:C5-C6}
\end{align}
In $1+1$ dimensions the scalar field is dimensionless and $\lambda$ has mass dimension two.  At fixed $m$, the dimensionless loop parameter is $g=\lambda/m^2$. Our renormalization prescription is fixed by normal ordering at the
tree-level vacuum mass $m$; we do not impose an additional finite
on-shell counterterm.

For later use, we now write the Schr\"odinger-picture fields in terms
of vacuum plane-wave oscillators,
\begin{align}
\phi(x)
&=
\int\frac{\dd p}{2\pi}
\frac{a_p^\dagger+a_{-p}}{\sqrt{2\omega_p}}
e^{-ipx},
\label{eq:plane-wave-phi}
\\
\pi(x)
&=
i\int\frac{\dd p}{2\pi}
\sqrt{\frac{\omega_p}{2}}
\left(a_p^\dagger-a_{-p}\right)e^{-ipx},
\label{eq:plane-wave-pi}
\\
\omega_p
&=\sqrt{m^2+p^2}.
\label{eq:plane-wave-dispersion}
\end{align}
Here $p$ denotes the vacuum spatial momentum and $\omega_p$ its
frequency.  The normalization is chosen such that
\begin{equation}
[\phi(x),\pi(y)]=i\delta(x-y),
\qquad
[a_p,a_q^\dagger]=2\pi\delta(p-q).
\label{eq:canonical-algebra}
\end{equation}
With this convention, normal ordering in
Eqs.~\eqref{eq:H0-density} and \eqref{eq:H1-density} simply means that
all creation operators $a^\dagger$ are placed to the left of all
annihilation operators $a$.  The explicit cubic and higher terms in
$\mathcal H_1$ do not enter the one-loop spectrum directly; their
contributions begin at higher orders in the semiclassical expansion.

\subsection{Kink sector and quadratic fluctuation operator}
\label{subsec:kink-sector}
We now move from the perturbative vacuum to the one-kink sector.  In
terms of the shifted field, the classical background is
\begin{equation}
f(x)\equiv\Phi_K(x)+v
=
v\left[
1+
\frac{\epsilon\sinh(mx/2)}
{\sqrt{1+\epsilon^2\cosh^2(mx/2)}}
\right].
\label{eq:shifted-kink}
\end{equation}
The shift is useful because the left vacuum now corresponds to
$f(-\infty)=0$, while the kink approaches $f(+\infty)=2v$ on the
right. Let $|K\rangle$ denote the one-kink state.  We define its energy and
mass relative to the perturbative vacuum by
\begin{equation}
H|K\rangle=E_K|K\rangle,
\qquad
M_K=E_K-E_0.
\label{eq:kink-eigenvalue-mass}
\end{equation}
The centered profile $f(x)$ gives the leading semiclassical background.
Strictly speaking, a localized quantum kink requires a wave packet in
the collective coordinate, since an exact momentum eigenstate cannot
also be a normalizable state centered at a fixed position
\cite{zerostates,moving}.

Following the operator construction of Ref.~\cite{evslin2019}, we
introduce the displacement operator
\begin{equation}
\mathcal D_f
=
\exp\!\left[-i\int\dd x\,f(x)\pi(x)\right].
\label{eq:displacement-operator}
\end{equation}
Because the kink profile is not square-integrable, this expression
should be understood as a regulated shift between topological sectors,
rather than as a globally unitary operator acting on the infinite-volume
vacuum Fock space.  What we shall need is its local action,
\begin{equation}
\mathcal D_f^\dagger\phi(x)\mathcal D_f
=
\phi(x)+f(x),
\qquad
\mathcal D_f^\dagger\pi(x)\mathcal D_f
=
\pi(x).
\label{eq:displacement-action}
\end{equation}
Using this shift, one can move $\mathcal D_f$ through the Hamiltonian
and organize the result according to powers of the fluctuation field
\begin{equation}
H\mathcal D_f
=
\mathcal D_fH',
\qquad
H'
=
M_{\rm cl}+H_K+H_{\rm int}.
\label{eq:shifted-Hamiltonian}
\end{equation}
The first term is simply the classical kink energy already obtained in
Eq.~\eqref{eq:classical}.  In the shifted notation it may equivalently
be written as
\begin{equation}
M_{\rm cl}
=
\int\dd x\,
\left[
\frac12(f')^2
+
\frac{m^2}{2}f^2
+C_3f^3+C_4f^4+C_5f^5+C_6f^6
\right].
\label{eq:classical-kink-mass}
\end{equation}
The term linear in the quantum fluctuation vanishes because $f(x)$
satisfies the static classical equation of motion.  The first
nontrivial contribution is therefore quadratic,
\begin{equation}
H_K
=
\frac12
\int\dd x\,
\left[
:\pi^2:
+:(\partial_x\phi)^2:
+\mathcal V''(f(x)):\phi^2:
\right].
\label{eq:quadratic-Hamiltonian}
\end{equation}
This is the Hamiltonian that determines the one-loop fluctuation
spectrum.

To separate the asymptotic vacuum mass from the localized distortion
generated by the kink, we write
\begin{equation}
\mathcal V''(f(x))=m^2-U_\epsilon(x),
\qquad
U_\epsilon(x)=m^2u_\epsilon(X),
\label{eq:U-definition}
\end{equation}
where $X=mx$ is the dimensionless coordinate introduced in
Eq.~\eqref{eq:dimensionless-classical-variables}.  In terms of the
normalized kink profile $\chi_\epsilon(X)$, the dimensionless localized
potential is
\begin{equation}
u_\epsilon(X)
=
\frac{3\left[1-\chi_\epsilon^2(X)\right]
\left[1+2\epsilon^2+5\chi_\epsilon^2(X)\right]}
{4(1+\epsilon^2)}.
\label{eq:localized-potential}
\end{equation}
Since $u_\epsilon(X)$ vanishes at large $|X|$, we recover
$\mathcal V''\to m^2$ asymptotically, as expected from the vacuum mass
defined in Eq.~\eqref{eq:vacuum-mass}.

For the spectral analysis it is convenient to introduce the
dimensionless Hessian potential and operator,
\begin{equation}
W_\epsilon(X)
\equiv
\frac{\mathcal V''(f(x))}{m^2}
=1-u_\epsilon(X),
\qquad
\mathcal K_\epsilon
\equiv
-\partial_X^2+W_\epsilon(X).
\label{eq:dimensionless-hessian}
\end{equation}
The corresponding dimensional fluctuation operator is
\begin{equation}
\widehat{\mathcal K}_\epsilon
\equiv
m^2\mathcal K_\epsilon
=
-\partial_x^2+\mathcal V''(f(x)).
\label{eq:dimensional-hessian}
\end{equation}
These definitions will be used throughout the spectral and one-loop sections below.

For completeness, the terms of cubic and higher order generated by
the shift are
\begin{equation}
H_{\rm int}
=
\int\dd x\,
\left[
\Gamma_3(x):\phi^3:
+\Gamma_4(x):\phi^4:
+\Gamma_5(x):\phi^5:
+\Gamma_6:\phi^6:
\right],
\label{eq:shifted-interactions}
\end{equation}
with
\begin{align}
\Gamma_3(x)
&=
C_3+4C_4f+10C_5f^2+20C_6f^3,
\label{eq:Gamma3}
\\
\Gamma_4(x)
&=
C_4+5C_5f+15C_6f^2,
\label{eq:Gamma4}
\\
\Gamma_5(x)
&=
C_5+6C_6f,
\qquad
\Gamma_6=C_6.
\label{eq:Gamma5-Gamma6}
\end{align}
At one loop only the quadratic Hamiltonian in Eq.~\eqref{eq:quadratic-Hamiltonian} and the normal-ordering subtraction are required while the explicit interaction Hamiltonian $H_{\rm int}$ starts contributing at higher orders.

% ============================================================

% ============================================================
\section{Fluctuation spectrum in general-Heun form}
\label{sec:heun}

\subsection{Reduction to the Heun equation}
\label{subsec:heun-reduction}

To analyze the fluctuation spectrum analytically, we now recast the mode equation in general-Heun form.  From this point onward we assume $\epsilon>0$.  The reason is simply that the change of variables introduces
$a=(1+\epsilon^2)/\epsilon^2$, which is singular at $\epsilon=0$.  The strict $\epsilon=0$ theory should therefore not be obtained by inserting $\epsilon=0$ into the formulas below.  Instead, its relation to the finite-$\epsilon$ family is understood through the $\epsilon\to0^+$ limits discussed in Sec.~\ref{subsubsec:two-subkink} and Sec.~\ref{subsec:small-spectrum}.

Let $g_k(x)$ denote a continuum fluctuation of the dimensional operator $\widehat{\mathcal K}_\epsilon$ introduced in Eq.~\eqref{eq:dimensional-hessian}.  We write its eigenvalue as
\begin{equation}
\widehat{\mathcal K}_\epsilon g_k(x)
=
\omega_k^2 g_k(x),
\qquad
\omega_k^2=m^2+k^2,
\label{eq:fluctuation-schrodinger}
\end{equation}
where $k\in\mathbb R$ is the continuum momentum measured relative to the asymptotic vacuum.  Using the definition of $U_\epsilon$ in Eq.~\eqref{eq:U-definition}, the same equation takes the more convenient Schr\"odinger form
\begin{equation}
g_k''(x)+\left[k^2+U_\epsilon(x)\right]g_k(x)=0.
\label{eq:fluctuation-equation}
\end{equation}

It is useful to introduce the variables
\begin{equation}
z=-\sinh^2\!\left(\frac{mx}{2}\right),
\qquad
D_\epsilon(x)=1+\epsilon^2\cosh^2\!\left(\frac{mx}{2}\right),
\qquad
a=\frac{1+\epsilon^2}{\epsilon^2},
\qquad
\nu=\frac{k}{m}.
\label{eq:heun-variables}
\end{equation}
With these definitions $D_\epsilon=\epsilon^2(a-z)$.  Writing $g_k(x)=G_k(z)$ and carrying out the change of variable, one finds
\begin{equation}
z(z-1)G_k''
+
\frac{2z-1}{2}G_k'
+
\left[
\nu^2
+
\frac{3}{4}
\frac{(1+\epsilon^2)(1+2\epsilon^2)-2\epsilon^2(\epsilon^2+3)z}
{(1+\epsilon^2-\epsilon^2z)^2}
\right]G_k
=0.
\label{eq:G-equation}
\end{equation}

The additional singular point appears at $z=a$, where the coefficient of $G_k$ has a double pole.  A local Frobenius analysis gives the two indicial exponents $-3/2$ and $5/2$.  We factor out the $-3/2$ branch,
\begin{equation}
G_k(z)
=
\left(\frac{a}{a-z}\right)^{3/2}H_k(z),
\label{eq:heun-factorization}
\end{equation}
which leaves an equation with the standard four regular singular points of the general Heun problem:
\begin{equation}
H_k''
+
\left[
\frac{1}{2z}
+
\frac{1}{2(z-1)}
-
\frac{3}{z-a}
\right]H_k'
+
\frac{(\nu^2+9/4)z-a(\nu^2+3/4)}
{z(z-1)(z-a)}H_k
=0.
\label{eq:heun-reduced}
\end{equation}

For definiteness, we use the convention in which $\HeunG(a,q;\alpha,\beta,\gamma,\delta;z)$ solves
\begin{equation}
Y''
+
\left[
\frac{\gamma}{z}
+
\frac{\delta}{z-1}
+
\frac{\eta}{z-a}
\right]Y'
+
\frac{\alpha\beta z-q}{z(z-1)(z-a)}Y
=0,
\qquad
\eta=\alpha+\beta-\gamma-\delta+1.
\label{eq:canonical-heun}
\end{equation}
Matching Eq.~\eqref{eq:heun-reduced} term by term with this canonical form gives
\begin{equation}
\alpha=-\frac{3}{2}+i\nu,
\qquad
\beta=-\frac{3}{2}-i\nu,
\qquad
\gamma=\delta=\frac{1}{2},
\qquad
q=a\left(\nu^2+\frac{3}{4}\right).
\label{eq:heun-parameters-even}
\end{equation}
These parameters will be used below to construct solutions of definite parity.

\subsection{Even and odd continuum modes}
\label{subsec:parity-modes}

Because the fluctuation potential is even in $x$, the continuum solutions can be chosen with definite parity.  The point $x=0$, corresponding to $z=0$, is then the natural place to fix their local normalization.  In the even sector we take the regular Frobenius solution and choose $E_k(0)=1$; parity automatically gives $E_k'(0)=0$.  This leads to
\begin{align}
E_k(x)
={}&
\left[\frac{1+\epsilon^2}{D_\epsilon(x)}\right]^{3/2}
\HeunG\!\Bigg(
 a,
a\left(\nu^2+\frac{3}{4}\right);
-\frac{3}{2}+i\nu,
-\frac{3}{2}-i\nu,
\frac{1}{2},
\frac{1}{2};
-\sinh^2\!\frac{mx}{2}
\Bigg),
\label{eq:even-continuum}
\\
E_k(0)
={}&1,
\qquad
E_k'(0)=0.
\label{eq:even-origin}
\end{align}

For the odd sector one uses the second Frobenius branch at the origin.  It is convenient to fix its overall factor by requiring a unit slope, $O_k'(0)=1$, while odd parity gives $O_k(0)=0$.  With this choice one obtains
\begin{align}
O_k(x)
={}&
\frac{2}{m}
\left[\frac{1+\epsilon^2}{D_\epsilon(x)}\right]^{3/2}
\sinh\!\left(\frac{mx}{2}\right)\HeunG\!\Bigg(
 a,
a(\nu^2+1)-\frac{3}{2};
-1+i\nu,
-1-i\nu,
\frac{3}{2},
\frac{1}{2};
-\sinh^2\!\frac{mx}{2}
\Bigg),
\label{eq:odd-continuum}
\\
O_k(0)
={}&0,
\qquad
O_k'(0)=1.
\label{eq:odd-origin}
\end{align}

Thus Eqs.~\eqref{eq:even-continuum} and \eqref{eq:odd-continuum} give exact even and odd solutions of Eq.~\eqref{eq:fluctuation-equation}.  A direct substitution check is collected in Appendix~\ref{app:verification}.  It is worth stressing, however, that the normalization used here is only a convenient normalization at the origin.  For continuum quantization we still have to rescale these functions by momentum-dependent factors, chosen from their large-$|x|$ behavior, so that they satisfy the usual scattering normalization.  Keeping these two normalizations distinct will be useful when we turn to the continuum spectral density and the one-loop energy.

% ============================================================
\section{Bound states and numerical spectrum}
\label{sec:bound-states}

\subsection{Heun quantization conditions}
\label{subsec:bound-quantization}
We now turn from continuum scattering states to the discrete spectrum. Since the continuum starts at $\omega=m$, a bound state corresponds to
an imaginary continuum momentum.  It is therefore convenient to write
\begin{equation}
k=i\kappa,
\qquad
\kappa=m\sigma,
\qquad
0<\sigma\leq1,
\qquad
\omega_\sigma=m\sqrt{1-\sigma^2}.
\label{eq:bound-parametrization}
\end{equation}
In this parametrization, $\sigma=0$ corresponds to the continuum
threshold, whereas $\sigma=1$ gives a zero-frequency state. The bound-state solutions follow directly by analytically continuing
the parity-resolved Heun functions obtained in Sec.~\ref{sec:heun}.
For the even sector, substituting $\nu=i\sigma$ into
Eq.~\eqref{eq:even-continuum} gives
\begin{equation}
E_\sigma(x)
=
\left[\frac{1+\epsilon^2}{D_\epsilon(x)}\right]^{3/2}
\HeunG\!\left(
 a,
a\left(\frac{3}{4}-\sigma^2\right);
-\frac{3}{2}-\sigma,
-\frac{3}{2}+\sigma,
\frac{1}{2},
\frac{1}{2};
-\sinh^2\!\frac{mx}{2}
\right),
\label{eq:even-bound-mode}
\end{equation}
while the corresponding odd solution is
\begin{align}
O_\sigma(x)
={}&
\frac{2}{m}
\left[\frac{1+\epsilon^2}{D_\epsilon(x)}\right]^{3/2}
\sinh\!\left(\frac{mx}{2}\right)\HeunG\!\left(
 a,
a(1-\sigma^2)-\frac{3}{2};
-1-\sigma,
-1+\sigma,
\frac{3}{2},
\frac{1}{2};
-\sinh^2\!\frac{mx}{2}
\right).
\label{eq:odd-bound-mode}
\end{align}
To turn these local parity solutions into a quantization condition, we
examine their behavior at large positive $x$.  Let
$j_\pm(x)$ denote the exact Jost branches, normalized asymptotically by
\begin{equation}
  j_\pm(x)=e^{\pm m\sigma x}[1+o(1)].  
\end{equation}
The even and odd solutions can then be written as
\begin{align}
E_\sigma(x)
&=
A_e(\sigma,\epsilon)j_+(x)
+B_e(\sigma,\epsilon)j_-(x),
\label{eq:even-bound-asymptotic}
\\
O_\sigma(x)
&=
A_o(\sigma,\epsilon)j_+(x)
+B_o(\sigma,\epsilon)j_-(x).
\label{eq:odd-bound-asymptotic}
\end{align}
A normalizable state cannot contain the exponentially growing branch.
The discrete spectrum is therefore selected by
\begin{equation}
A_e(\sigma,\epsilon)=0
\quad\text{or}\quad
A_o(\sigma,\epsilon)=0,
\qquad
0<\sigma\leq1.
\label{eq:heun-quantization}
\end{equation}
Thus the zeros of the two growing-branch coefficients give the even
and odd bound-state spectra, respectively.

An explicit global connection formula for the Heun function is not
needed in order to determine these coefficients.  For example, the
even coefficient may be extracted directly from the asymptotic
solution as
\begin{equation}
A_e(\sigma,\epsilon)
=
\lim_{L\to\infty}
\frac{e^{-m\sigma L}}{2m\sigma}
\left[E_\sigma'(L)+m\sigma E_\sigma(L)\right],
\label{eq:Ae-extraction}
\end{equation}
and the same expression with $E_\sigma$ replaced by $O_\sigma$ defines
$A_o$.  In this sense, Eq.~\eqref{eq:Ae-extraction} plays the same role
as reading off the coefficient of the growing exponential in the
modified P\"oschl--Teller problem.

It is useful to clarify the respective roles of the Heun representation
and the numerical calculation used below.  The formulas above give an
exact analytic parametrization of the parity solutions and reduce the
bound-state problem to the implicit conditions in
Eq.~\eqref{eq:heun-quantization}.  Numerically, however, we do not
attempt to evaluate a global Heun connection formula directly.
Instead, we integrate the same fluctuation equation using asymptotic
boundary data and locate the corresponding roots by shooting.  The
Heun reduction therefore organizes the exact spectral problem, while
the shooting calculation provides a practical and independent way of
extracting its discrete eigenvalues.

\subsection{Exact zero mode and linear stability}
\label{subsec:zero-mode}

Before discussing the numerical spectrum, one bound state can be
identified without any numerical calculation.  Translational
invariance requires the derivative of the kink profile to be a zero
mode.  Indeed, the static background satisfies
\begin{equation}
-f''(x)+\mathcal V'(f(x))=0,
\label{eq:static-kink-eom}
\end{equation}
and differentiating this equation with respect to $x$ gives
\begin{equation}
\left[-\frac{\dd^2}{\dd x^2}+\mathcal V''(f(x))\right]f'(x)=0.
\label{eq:zero-mode-operator}
\end{equation}
Hence the translational mode has $\omega_0=0$.  In the parametrization
introduced above this corresponds to
\begin{equation}
\sigma_0=1,
\qquad
\kappa_0=m,
\qquad
k_0=im.
\label{eq:zero-mode-momentum}
\end{equation}
For later use, we choose the zero mode to have unit value at the
origin.  This gives
\begin{equation}
g_0(x)
=
\frac{(1+\epsilon^2)^{3/2}\cosh(mx/2)}
{\left[1+\epsilon^2\cosh^2(mx/2)\right]^{3/2}}.
\label{eq:zero-mode}
\end{equation}
Its norm is defined by
\begin{equation}
C_0^2
=
\int_{-\infty}^{\infty}\dd x\,g_0^2(x)
\label{eq:C0-definition}
\end{equation}
and in this case the integral can be carried out analytically
\begin{equation}
C_0^2
=
\frac{1}{2m}
\left[
\frac{(1+\epsilon^2)(2-\epsilon^2)}{\epsilon^2}
+
\sqrt{1+\epsilon^2}(\epsilon^2+4)
\operatorname{artanh}\!\left(\frac{1}{\sqrt{1+\epsilon^2}}\right)
\right].
\label{eq:C0-analytic}
\end{equation}
The zero mode also gives a simple stability check.  The function
$g_0(x)$ is strictly positive and therefore has no nodes.  By the
Sturm oscillation theorem it must be the ground state of the
one-dimensional Schr\"odinger operator.  There can consequently be no state below it with $\omega^2<0$, which establishes classical linear stability for every $\epsilon>0$.  The same conclusion can be expressed through the factorization $
\widehat{\mathcal K}_\epsilon=A^\dagger A,$ with $
A=\partial_x-\frac{g_0'}{g_0},$
as in the standard supersymmetric-quantum-mechanics construction
\cite{cooper}.  

\subsection{Numerical spectrum and threshold crossings}\label{subsec:numerical-bound-spectrum}
We next determine the remaining bound states numerically.  The shooting
calculation is performed on the dimensionless half-line
$0\leq X\leq L$, with $L=36$.  We use relative and absolute tolerances
$2\times10^{-10}$ and $2\times10^{-12}$, respectively.  The exact
translational mode is inserted analytically, while the remaining
eigenvalues are obtained from the appropriate even or odd boundary
conditions.  The resulting spectrum is summarized in
Table~\ref{tab:bound}. 

The numerical spectrum in Table~\ref{tab:bound} follows the expected
Sturm ordering, with even and odd states appearing alternately.
A second trend becomes clear for smaller $\epsilon$: additional
bound states enter below the continuum threshold.  This behavior is
consistent with the small-$\epsilon$ geometry discussed in
Sec.~\ref{subsubsec:two-subkink}, where the two interfaces move farther
apart and create an increasingly extended central region.  In the next
subsection we show how this widening region leads naturally to a
cavity-like spectrum and explains the growth in the number of bound
states, complementing the classical picture discussed in
Ref.~\cite{dorey2023}.
\begin{table}[H]
\centering
\small
\caption{Bound-state frequencies $\omega/m$. The count includes the zero
mode. Superscripts indicate even (E) or odd (O) parity with $N$ is the total number of bound states.}
\label{tab:bound}
\begin{tabular}{@{}crl@{}}
\toprule
$\eps$ & $N$ & $\omega/m$\\
\midrule
$0.05$&7&$0^{\rm E},\ 0.049799^{\rm O},\ 0.554826^{\rm E},\ 0.669831^{\rm O},$\\
&&$0.797614^{\rm E},\ 0.914441^{\rm O},\ 0.994733^{\rm E}$\\
$0.10$&6&$0^{\rm E},\ 0.098715^{\rm O},\ 0.585277^{\rm E},\ 0.732174^{\rm O},$\\
&&$0.875351^{\rm E},\ 0.981200^{\rm O}$\\
$0.20$&5&$0^{\rm E},\ 0.192141^{\rm O},\ 0.643954^{\rm E},\ 0.821732^{\rm O},\ 0.960342^{\rm E}$\\
$0.50$&4&$0^{\rm E},\ 0.422716^{\rm O},\ 0.795124^{\rm E},\ 0.962891^{\rm O}$\\
$1.00$&3&$0^{\rm E},\ 0.643421^{\rm O},\ 0.926923^{\rm E}$\\
$2.00$&3&$0^{\rm E},\ 0.789855^{\rm O},\ 0.988028^{\rm E}$\\
$5.00$&3&$0^{\rm E},\ 0.852457^{\rm O},\ 0.999534^{\rm E}$\\
$10.0$&3&$0^{\rm E},\ 0.862579^{\rm O},\ 0.99996877^{\rm E}$\\
$20.0$&3&$0^{\rm E},\ 0.865161^{\rm O},\ 0.99999801^{\rm E}$\\
$50.0$&3&$0^{\rm E},\ 0.865887^{\rm O},\ 0.999999949^{\rm E}$\\
\bottomrule
\end{tabular}
\end{table}

A new bound state appears when one of the fluctuation modes reaches the
continuum threshold $\omega=m$.  In terms of the bound-state parameter
introduced above, this corresponds to $\sigma=0$.  At this point the
large-$X$ solution no longer decays exponentially, but behaves as
\begin{equation}
g(X)\sim c_1+c_2X.
\end{equation}
For the solution to remain bounded as $X\to\infty$, the coefficient of
the growing term must vanish, $c_2=0$. The remaining constant solution is bounded but not square integrable,
so it represents a threshold state rather than a proper bound state.

We determine the values of $\epsilon$ for which this threshold
condition is satisfied by imposing the constant asymptotic branch at
large positive $X$ and integrating inward.  The first eight threshold
values are listed in Table~\ref{tab:thresholds}.

\begin{table}[H]
\centering
\caption{Threshold values $\epsilon_n$ at which a fluctuation mode reaches
the continuum threshold $\omega=m$.  The corresponding parity of the
threshold state is also shown.}
\label{tab:thresholds}
\begin{tabular}{ccc}
\toprule
$n$ & Parity & $\epsilon_n$ \\
\midrule
1 & O & 0.85559968 \\
2 & E & 0.35502899 \\
3 & O & 0.14979446 \\
4 & E & 0.06145327 \\
5 & O & 0.02491517 \\
6 & E & 0.01006926 \\
7 & O & 0.00406646 \\
8 & E & 0.00164199 \\
\bottomrule
\end{tabular}
\end{table}

\subsubsection{Exact first odd threshold and comparison of parameterizations}
\label{subsubsec:exact-threshold}

The first nontrivial crossing in Table~\ref{tab:thresholds} can in fact
be obtained analytically from the Heun representation, and this gives a
useful check on both the numerical shooting and the parameter
normalization.  At threshold, \(\sigma=0\), the odd solution in
Eq.~\eqref{eq:odd-bound-mode} contains
\begin{equation}
H(z)=\HeunG\!\left(a,a-\frac32;-1,-1,\frac32,\frac12;z\right),
\qquad
z=-\sinh^2\!\frac{mx}{2}.
\end{equation}
Because \(\alpha=\beta=-1\), a degree-one Heun polynomial is possible.
Substituting \(H(z)=1+bz\) into Eq.~\eqref{eq:canonical-heun} gives
\begin{equation}
3b(a-1)=1,
\qquad
3ab=2a-3.
\end{equation}
Eliminating \(b\) yields
\begin{equation}
2a^2-6a+3=0,
\qquad
a=\frac{3\pm\sqrt3}{2}.
\end{equation}
For the present family \(a=1+\epsilon^{-2}>1\), so only the plus sign is
admissible.  Consequently
\begin{equation}
\epsilon_1^2=\sqrt3-1,
\qquad
\epsilon_1=\sqrt{\sqrt3-1}=0.855599677167\ldots .
\label{eq:exact-eps1}
\end{equation}
The corresponding bounded, non-square-integrable odd threshold mode
may be written, up to an overall constant, as
\begin{equation}
O_{\rm th}(x)\propto
\left[\frac{1+\epsilon_1^2}{D_{\epsilon_1}(x)}\right]^{3/2}
\sinh\!\frac{mx}{2}
\left[1-\frac{\epsilon_1^2}{3}
\sinh^2\!\frac{mx}{2}\right].
\label{eq:exact-threshold-mode}
\end{equation}
Its approach to a finite nonzero constant at spatial infinity is the
expected threshold behavior.

This exact result also clarifies the relation to two parametrizations
used in the recent literature.  Ref.~\cite{dorey2023} uses
\(\beta=\log(1+1/\epsilon)\); Eq.~\eqref{eq:exact-eps1} therefore gives
\begin{equation}
\beta_1=\log\!\left(1+\frac{1}{\sqrt{\sqrt3-1}}\right)
=0.774160599\ldots .
\end{equation}
The value differs from the approximate location \(\beta\simeq0.73289\)
quoted for the corresponding new mode in Ref.~\cite{dorey2023}.  This
is not a mass- or coordinate-normalization effect: after setting the
outer-vacuum mass to two, the fluctuation operator in that reference is
the same operator as the one used here.  In the alternative
parametrization of Ref.~\cite{blaschke2026},
\begin{equation}
U_{\rm CL}=\frac12\frac{1+c\phi^2}{1+c}(\phi^2-1)^2,
\qquad c=\epsilon^{-2},
\end{equation}
our exact crossing becomes
\begin{equation}
c_3=\frac{1+\sqrt3}{2}=1.366025403\ldots ,
\end{equation}
in agreement with the independently reported numerical value
\(c_3=1.36603\)~\cite{blaschke2026}.  We therefore use
Eq.~\eqref{eq:exact-eps1} as an analytic anchor and determine the higher
crossings by the same bounded-threshold shooting condition used for
Table~\ref{tab:thresholds}.

Each value $\epsilon_n$ marks the point at which one mode sits exactly
at the continuum threshold.  For $\epsilon$ slightly larger than
$\epsilon_n$, the mode still belongs to the continuum.  At
$\epsilon=\epsilon_n$, it becomes a bounded threshold state with
$\omega=m$.  For slightly smaller $\epsilon$, the same mode moves below
the threshold and becomes a normalizable bound state.  The successive
states alternate between odd and even parity, in agreement with the
usual ordering of eigenstates in a one-dimensional Schr\"odinger
problem.

The threshold state itself is not included in the bound-state count
because it is not square integrable.  We also checked the numerical
stability of the values in Table~\ref{tab:thresholds} by moving the
integration endpoint farther outward.  Once the endpoint lies well
beyond the two interfaces, the displayed values remain unchanged to
the quoted precision. A threshold crossing does not create a new contribution to the
spectrum from nothing.  Instead, the same mode changes its character.
Before the crossing it contributes through the continuum spectrum,
whereas after the crossing it appears as a discrete bound state.
Consequently, the division between the discrete and continuum parts of
the spectrum changes at $\epsilon_n$, while the complete regulated
spectral trace remains continuous.  A similar transfer of spectral
weight occurs in spectral-wall problems~\cite{walls}, although in the
present case the crossing is controlled by the deformation parameter
$\epsilon$ rather than by a position in moduli space.

\begingroup
\subsection{Small-\texorpdfstring{$\epsilon$}{epsilon} spectral anatomy:
cavity modes and threshold hierarchy}
\label{subsec:small-spectrum}

The growth of the bound-state count at small $\epsilon$ can be
understood quite directly from the two-subkink geometry.  Recall that
the dimensionless Hessian potential is
$W_\epsilon(X)=\mathcal V''(f)/m^2$.  When
$\epsilon\to0^+$, the separation between the two interfaces becomes
large and $W_\epsilon$ develops two different asymptotic plateaux,
\begin{equation}
W_\eps(X)\longrightarrow
\begin{cases}
1/4, & |X|\ll \ell_X/2,\\[1mm]
1, & |X|\gg \ell_X/2,
\end{cases}
\qquad
\ell_X=4\log(2/\eps)+O(1).
\label{eq:cavity-plateaux}
\end{equation}
The origin of the additional bound states can be seen directly from
this shape of the fluctuation potential.  Figure~\ref{fig:fluctuation-cavity}
shows $W_\epsilon(X)$ for several values of $\epsilon$.  For small
$\epsilon$, the potential contains a broad central region with
$W_\epsilon\simeq1/4$, bounded on both sides by the two separated
interfaces.  Far outside this region, the potential approaches the
true-vacuum value $W_\epsilon=1$.  As $\epsilon$ becomes smaller, the
central region becomes wider while its limiting height remains close
to $1/4$.  The fluctuation problem therefore takes the form of an
increasingly long effective cavity.

\begin{figure}[H]
\centering
\includegraphics[width=.82\textwidth]{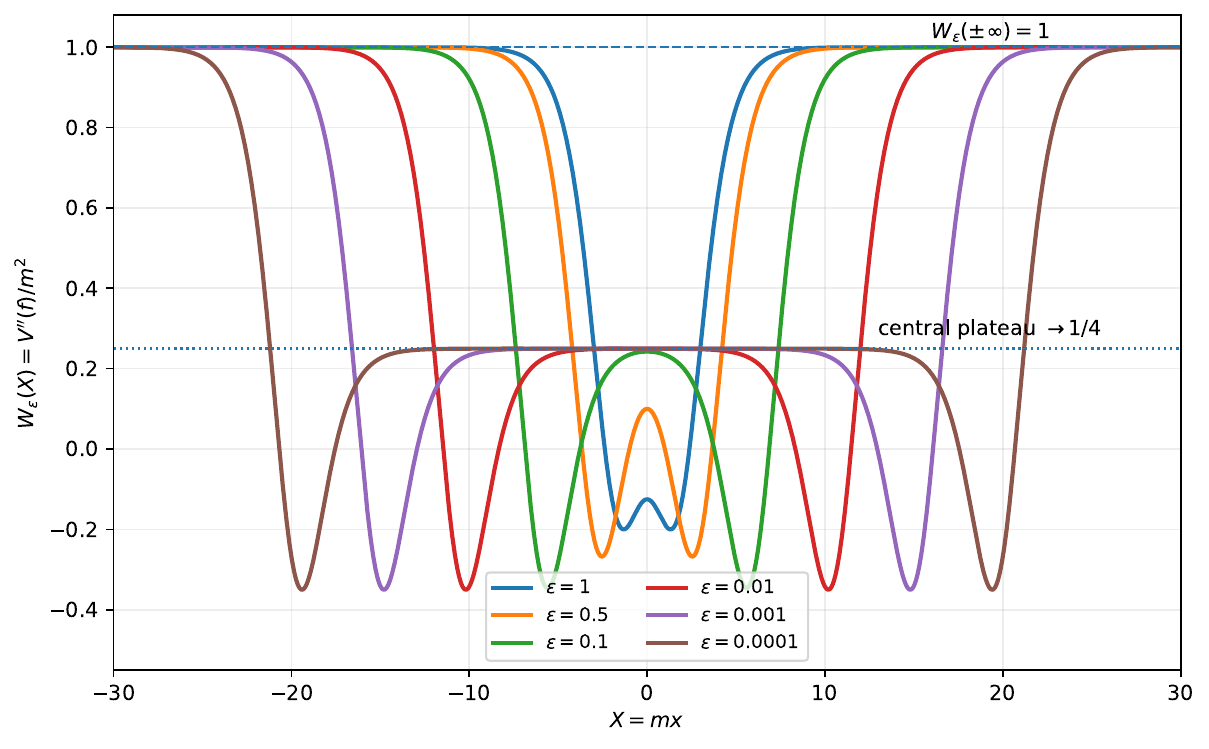}
\caption{Dimensionless fluctuation potential
$W_\epsilon(X)=\mathcal V''(f)/m^2$ for several values of the
deformation parameter $\epsilon$.  For small $\epsilon$, the region
between the two interfaces approaches the plateau
$W_\epsilon=1/4$, while the outer regions approach
$W_\epsilon=1$.  As $\epsilon$ becomes smaller, the two interfaces
move farther apart and the central plateau becomes wider.  This
extended region acts as an effective cavity for fluctuation modes with
$1/2<\omega/m<1$.}
\label{fig:fluctuation-cavity}
\end{figure}

This picture suggests a simple interpretation of the low-lying
spectrum.  Modes with frequencies between the central and outer
continuum scales can oscillate inside the central region but decay
outside it.  To describe these states, we introduce the dimensionless
frequency
\begin{equation}
\varpi\equiv\frac{\omega}{m}.
\end{equation}
Consider first a mode in the range
\begin{equation}
\frac12<\varpi<1.
\end{equation}
Inside the central plateau such a mode oscillates with dimensionless
wavenumber
\begin{equation}
\rho\equiv\sqrt{\varpi^2-\frac14},
\label{eq:cavity-q}
\end{equation}
whereas in the outer true-vacuum regions it decays with exponent
\begin{equation}
\kappa_{\rm out}\equiv\sqrt{1-\varpi^2}.
\end{equation}
We use the symbol $\rho$ rather than $q$ here in order to avoid
confusion with the Heun accessory parameter introduced in
Sec.~\ref{sec:heun}.

Each interface contributes an order-one reflection phase.  If
$\delta(\rho)$ denotes the corresponding phase for a single interface,
then the standing-wave condition across the long central region takes
the generic form
\begin{equation}
\rho\ell_X+2\delta(\rho)=n\pi+o(1),
\qquad
\eps\to0^+.
\label{eq:cavity-quantization}
\end{equation}
To leading order, the spacing in $\rho$ is therefore
$\Delta\rho\simeq\pi/\ell_X$. At the outer continuum threshold $\varpi=1$, the largest central
wavenumber is
\begin{equation}
\rho_*=\sqrt{1-\frac14}=\frac{\sqrt3}{2}.
\end{equation}
One then obtains the leading estimate
\begin{equation}
\displaystyle
N_{\rm cav}(\eps)
=\frac{\rho_*\ell_X}{\pi}+O(1)
=\frac{2\sqrt3}{\pi}\log\frac{2}{\eps}+O(1).
\label{eq:cavity-count}
\end{equation}
Thus the number of cavity states grows logarithmically as
$\epsilon\to0^+$.

Equation~\eqref{eq:cavity-count} should be interpreted as a counting
law rather than as a precision formula for individual eigenvalues.
It counts states in the band
$1/2<\omega/m<1$ and does not include either the exact translational
zero mode or the soft relative-translation mode discussed below.
For example, at $\epsilon=0.2,0.1,0.05$,
Table~\ref{tab:bound} contains $3,4,5$ cavity states, whereas
Eq.~\eqref{eq:cavity-count} gives $2.54,3.30,4.07$.  The difference is
of order unity, exactly where the reflection phase
$\delta(\rho)$ enters.  A quantitative prediction for each individual
level would require this phase to be determined; we therefore do not
assign more accuracy to Eq.~\eqref{eq:cavity-count} than its explicit
$O(1)$ remainder allows.

The cavity interpretation also provides a simple explanation for the
sequence of threshold crossings listed in
Table~\ref{tab:thresholds}.  At a threshold crossing,
$\rho=\rho_*=\sqrt3/2$.  To add one further standing-wave node, the
cavity length must increase, to leading order, by an amount satisfying
\begin{equation}
\rho_*\bigl(\ell_{X,n+1}-\ell_{X,n}\bigr)
=
\pi+o(1).
\end{equation}
Using
$\ell_X=4\log(2/\eps)+O(1)$, one finds the asymptotic geometric law
\begin{equation}
\displaystyle
\frac{\eps_{n+1}}{\eps_n}
\longrightarrow
\exp\!\left(-\frac{\pi}{2\sqrt3}\right)
=0.403774\ldots.
\label{eq:threshold-geometric}
\end{equation}
The seven successive ratios obtained from the numerical thresholds in
Table~\ref{tab:thresholds} are 0.41495, 0.42192, 0.41025, 0.40543, 0.40414, 0.40385, and 0.40379. They approach the predicted value 0.403774 rather rapidly.
Equivalently, $\log\epsilon_n$ becomes asymptotically linear in the
threshold index $n$, with slope $-\pi/(2\sqrt3)$.  The agreement over
this extended sequence is more informative than a comparison based on
only the first few crossings and gives strong numerical support to the
cavity interpretation of the alternating even and odd threshold
states.

There is, however, one low-frequency mode that does not belong to this
cavity tower.  In the strict $\epsilon=0$ theory, the two separated
elementary interfaces each carry their own translational zero mode.
For small positive $\epsilon$, the symmetric combination becomes the
exact translational mode of the composite kink, while the
antisymmetric combination corresponds to a relative displacement of
the two interfaces.

The tail of an isolated translation mode decays into the central
$\Phi\simeq0$ region with mass $m/2$.  The overlap of the two tails
across a separation $\ell_X$ is therefore of order
$e^{-\ell_X/2}$.  Using the logarithmic separation found in
Eq.~\eqref{eq:subkink-separation}, we obtain
\begin{equation}
\frac{\omega_{\rm rel}^2}{m^2}
=
O(e^{-\ell_X/2})
=
O(\epsilon^2),
\qquad
\text{hence}
\qquad
\frac{\omega_{\rm rel}}{m}=O(\epsilon).
\label{eq:relative-mode-overlap}
\end{equation}
This argument determines the scaling of the soft mode but not its
coefficient. The numerical spectrum is consistent with this expectation:
\begin{equation}
\frac{\omega_{\rm rel}}{m\epsilon}
=
0.9607\quad(\epsilon=0.2),
\qquad
0.9872\quad(\epsilon=0.1),
\qquad
0.9960\quad(\epsilon=0.05).
\end{equation}
The ratio clearly approaches a value close to unity.  We therefore use
$\omega_{\rm rel}/m\sim\epsilon$ as an asymptotic scaling supported by
both the overlap argument and the numerical data.  Without a full
collective-coordinate calculation, however, we do not interpret these
numbers as proving an exactly unit prefactor.  This mode should also be
distinguished from the cavity states: its frequency tends to zero,
whereas the cavity tower accumulates above $\omega=m/2$.

\endgroup

\subsection{The shallow even mode at large deformation}
\label{sec:shallow}

The opposite limit, $\epsilon\to\infty$, has a different spectral
feature.  In this regime the fluctuation operator approaches
\begin{equation}
\mathcal K_\infty
=
-\partial_X^2
+
1-\frac32\sech^2(X/2).
\end{equation}
This is the familiar $\phi^4$ fluctuation operator in the present
normalization.  It has two normalizable bound states,
\[
\frac{\omega}{m}=0,
\qquad
\frac{\omega}{m}=\frac{\sqrt3}{2},
\]
together with an even resonance exactly at the continuum threshold,
\begin{equation}
h_*(X)
=
\frac{3\tanh^2(X/2)-1}{2},
\qquad
h_*(\pm\infty)=1.
\end{equation}

For large but finite $\epsilon$, this threshold resonance moves slightly
below the continuum and becomes a very shallow bound state.  To see
this analytically, we introduce
\[
d=\eps^{-2},
\qquad
t=\tanh(X/2),
\]
and expand the localized well at fixed $X$:
\begin{equation}
u_\eps
=
\frac32(1-t^2)
+
\frac34d(1-t^2)(7t^2-1)
+
O(d^2).
\end{equation}
Accordingly,
\[
\mathcal K_\eps
=
\mathcal K_\infty
+
dW_1
+
O(d^2),
\]
with
\[
W_1
=
-\frac34(1-t^2)(7t^2-1).
\]
Let $\sigma_2>0$ denote the binding exponent of the even state that
emerges from the $\phi^4$ threshold resonance.  Its frequency can then
be written as
\[
\frac{\omega_2}{m}
=
\sqrt{1-\sigma_2^2}.
\]
For a state born from a threshold resonance, weak-binding perturbation
theory \cite{simon}, or equivalently the integrated Wronskian identity,
gives
\begin{equation}
\sigma_2
=
-\frac d2
\int_{\mathbb R}
W_1h_*^2\,\dd X
+
O(d^2).
\end{equation}
At leading order, the two asymptotic tails contribute the boundary term
$2\sigma_2$, while the perturbation contributes
$d\int W_1h_*^2$.

The remaining integral is elementary after using
$\dd X=2\dd t/(1-t^2)$:
\begin{equation}
\int_{\mathbb R}W_1h_*^2\,\dd X
=
-\frac38
\int_{-1}^1
(7t^2-1)(3t^2-1)^2\,\dd t
=
-\frac85.
\end{equation}
We therefore obtain
\begin{equation}
\displaystyle
\sigma_2
=
\frac4{5\eps^2}
+
O(\eps^{-4}),
\qquad
1-\frac{\omega_2}{m}
=
\frac8{25\eps^4}
+
O(\eps^{-6}).
\label{eq:shallow}
\end{equation}
The positive sign of $\sigma_2$ shows that the threshold resonance is
indeed pulled below the continuum for sufficiently large but finite
$\epsilon$.  The numerical spectrum follows the predicted scaling:
$\eps^2\sigma_2$ is $0.79034$, $0.79756$, and $0.79961$ for
$\eps=10,20,50$, respectively, approaching the asymptotic value
$4/5$. This state is extremely shallow.  Its decay length grows as $\frac{5\eps^2}{4m}$, so a conventional finite-box eigenvalue calculation would require a
rapidly increasing spatial domain as $\epsilon$ grows.  Inward shooting
from the known asymptotic tail avoids this numerical difficulty.

% ============================================================
\section{One-loop kink energy: spectral formulation and deformation dependence}
\label{sec:one-loop}
\label{sec:results}

Having determined the bound and continuum spectra, we now turn to the
one-loop correction to the kink energy.  The calculation involves two
different but closely related bases.  The quadratic kink Hamiltonian is
naturally diagonalized in eigenmodes of
$\widehat{\mathcal K}_\epsilon$, whereas normal ordering was defined in
Sec.~\ref{subsec:vacuum-quantization} with respect to the vacuum
plane-wave operators $a_p$ and $a_p^\dagger$.  We therefore begin by
constructing a properly normalized spectral basis and relating the two
sets of oscillators.  This leads to the renormalized one-loop energy,
which we then evaluate numerically and study as a function of the
deformation parameter.

\subsection{Spectral basis}
\label{subsec:spectral-basis}

The Heun solutions obtained in Sec.~\ref{subsec:parity-modes} were
normalized at the origin.  That choice is useful for solving the mode
equation and separating the two parity sectors, but the continuum
spectral expansion requires the normalization to be fixed instead by
the large-$|x|$ scattering behavior.
For $k>0$, we write
\begin{equation}
E_k(x)\sim \mathcal N_e(k)\cos\!\bigl(kx+\delta_e(k)\bigr),
\qquad
O_k(x)\sim \mathcal N_o(k)\sin\!\bigl(kx+\delta_o(k)\bigr).
\end{equation}
Here $\mathcal N_e(k)$ and $\mathcal N_o(k)$ are the asymptotic
amplitudes, while $\delta_e(k)$ and $\delta_o(k)$ are the even and odd
phase shifts.  These quantities should not be confused with the
growing-branch coefficients $A_e(\sigma,\epsilon)$ and
$A_o(\sigma,\epsilon)$ used for bound-state quantization in
Sec.~\ref{subsec:bound-quantization}.  The latter determine whether a
bound-state solution contains an exponentially growing component,
whereas $\mathcal N_e$ and $\mathcal N_o$ fix the normalization of
oscillatory continuum states. We can therefore define the real scattering-normalized parity modes by
\begin{equation}
\psi_{e,k}=\frac{\sqrt2\,E_k}{\mathcal N_e(k)},
\qquad
\psi_{o,k}=\frac{\sqrt2\,O_k}{\mathcal N_o(k)}.
\end{equation}
With this convention,
\begin{equation}
\int\dd x\,
\psi_{s,k}(x)\psi_{s',k'}(x)
=
2\pi\delta_{ss'}\delta(k-k').
\end{equation}
It is convenient to combine the two parity channels into a complex
continuum basis,
\begin{equation}
\frac{g_k}{C_k}
=
\frac{\psi_{e,k}+i\psi_{o,k}}{\sqrt2},
\qquad
\frac{g_{-k}}{C_{-k}}
=
\left(\frac{g_k}{C_k}\right)^*,
\qquad
C_{-k}=C_k>0.
\end{equation}
The signed momentum label simply parametrizes two independent
continuum combinations; no assumption of reflectionless scattering is
being made.

This construction also makes clear why the origin-normalized Heun
functions cannot be inserted directly into the spectral expansion.
They must first be divided by their asymptotic scattering amplitudes.
In particular, the condition $O_k'(0)=1$ gives the unnormalized odd
solution a length dimension before this normalization is carried out. The relevant spectral equation is
\begin{equation}
\widehat{\mathcal K}_\epsilon g
=
\omega^2g,
\label{eq:K-operator}
\end{equation}
where $\widehat{\mathcal K}_\epsilon$ was introduced in
Eq.~\eqref{eq:dimensional-hessian}.  Its spectrum consists of continuum
states $g_k$, nonzero normalizable bound states $g_B$, and the
translational zero mode $g_0$.

For the continuum states we choose
\begin{equation}
g_k^*(x)=g_{-k}(x),
\qquad
\int\dd x\,
g_{k_1}^*(x)g_{k_2}(x)
=
2\pi C_{k_1}^2\delta(k_1-k_2),
\label{eq:continuum-normalization}
\end{equation}
and for each nonzero bound state we define
\begin{equation}
C_B^2
=
\int\dd x\,|g_B(x)|^2.
\label{eq:bound-normalization}
\end{equation}
The corresponding zero-mode normalization was already obtained in
Eqs.~\eqref{eq:C0-definition} and \eqref{eq:C0-analytic}.
With these conventions the three parts of the spectrum give the
completeness relation
\begin{equation}
\delta(x-y)
=
\int\frac{\dd k}{2\pi}
\frac{g_k(x)g_k^*(y)}{C_k^2}
+
\sum_B\frac{g_B(x)g_B^*(y)}{C_B^2}
+
\frac{g_0(x)g_0(y)}{C_0^2},
\label{eq:completeness}
\end{equation}
where the sum runs over all nonzero normalizable bound states of both
parities.

The fluctuation field can now be expanded directly in this basis:
\begin{align}
\phi(x)
={}&
\int\frac{\dd k}{2\pi}
\frac{g_k(x)}{C_k\sqrt{2\omega_k}}
\left(b_k^{\dagger}+b_{-k}\right)
+
\sum_B
\frac{g_B(x)}{C_B\sqrt{2\omega_B}}
\left(b_B^{\dagger}+b_B\right)
+
\phi_0\frac{g_0(x)}{C_0},
\label{eq:kink-mode-phi}
\\
\pi(x)
={}&
i\int\frac{\dd k}{2\pi}
\frac{g_k(x)}{C_k}
\sqrt{\frac{\omega_k}{2}}
\left(b_k^{\dagger}-b_{-k}\right)
+
i\sum_B
\frac{g_B(x)}{C_B}
\sqrt{\frac{\omega_B}{2}}
\left(b_B^{\dagger}-b_B\right)
+
\pi_0\frac{g_0(x)}{C_0}.
\label{eq:kink-mode-pi}
\end{align}
All bound-state eigenfunctions are chosen to be real, including the odd
states, so that the field operators remain Hermitian.  The canonical
commutation relations then imply
\begin{equation}
[b_{k_1},b_{k_2}^{\dagger}]
=
2\pi\delta(k_1-k_2),
\qquad
[b_B,b_{B'}^{\dagger}]
=
\delta_{BB'},
\qquad
[\phi_0,\pi_0]=i.
\label{eq:b-algebra}
\end{equation}
The pair $(\phi_0,\pi_0)$ describes the translational collective
coordinate rather than an ordinary oscillator.

\subsection{Transformation from plane-wave to kink oscillators}
\label{subsec:bogoliubov}

The need for two sets of oscillators is now apparent.  The vacuum
normal-ordering prescription is expressed in terms of $a_p$, whereas
the kink Hamiltonian is diagonal in the $b$ operators.  We therefore
need the transformation between the two descriptions. The inverse of the vacuum plane-wave expansion in
Eqs.~\eqref{eq:plane-wave-phi} and \eqref{eq:plane-wave-pi} is
\begin{align}
a_p^{\dagger}
&=
\int\dd x\,
\left[
\sqrt{\frac{\omega_p}{2}}\,\phi(x)
-
\frac{i}{\sqrt{2\omega_p}}\,\pi(x)
\right]e^{ipx},
\label{eq:apdag-inverse}
\\
a_{-p}
&=
\int\dd x\,
\left[
\sqrt{\frac{\omega_p}{2}}\,\phi(x)
+
\frac{i}{\sqrt{2\omega_p}}\,\pi(x)
\right]e^{ipx}.
\label{eq:aminus-inverse}
\end{align}
To perform the projection onto kink modes, we introduce
\begin{equation}
\widetilde g_k(p)
=
\int\dd x\,g_k(x)e^{ipx},
\qquad
\widetilde g_B(p)
=
\int\dd x\,g_B(x)e^{ipx},
\qquad
\widetilde g_0(p)
=
\int\dd x\,g_0(x)e^{ipx}.
\label{eq:mode-Fourier-transforms}
\end{equation}
Substituting Eqs.~\eqref{eq:kink-mode-phi} and
\eqref{eq:kink-mode-pi} into Eq.~\eqref{eq:apdag-inverse}, one finds
for the continuum part
\begin{equation}
a_{C,p}^{\dagger}
=
\int\frac{\dd k}{2\pi}
\frac{\widetilde g_k(p)}{2C_k}
\left[
\frac{\omega_p+\omega_k}{\sqrt{\omega_p\omega_k}}b_k^{\dagger}
+
\frac{\omega_p-\omega_k}{\sqrt{\omega_p\omega_k}}b_{-k}
\right],
\label{eq:continuum-bogoliubov}
\end{equation}
while a nonzero bound state contributes
\begin{equation}
a_{B,p}^{\dagger}
=
\frac{\widetilde g_B(p)}{2C_B}
\left[
\frac{\omega_p+\omega_B}{\sqrt{\omega_p\omega_B}}b_B^{\dagger}
+
\frac{\omega_p-\omega_B}{\sqrt{\omega_p\omega_B}}b_B
\right].
\label{eq:bound-bogoliubov}
\end{equation}
The zero mode must be treated separately because it is represented by
the collective-coordinate pair $(\phi_0,\pi_0)$ rather than by a
nonzero-frequency oscillator.  Its projection is
\begin{equation}
a_{0,p}^{\dagger}
=
\frac{\widetilde g_0(p)}{C_0}
\left[
\sqrt{\frac{\omega_p}{2}}\phi_0
-
\frac{i}{\sqrt{2\omega_p}}\pi_0
\right].
\label{eq:zero-bogoliubov}
\end{equation}
Equations~\eqref{eq:continuum-bogoliubov}--\eqref{eq:zero-bogoliubov}
are the Christ--Lee counterpart of the P\"oschl--Teller transformation
used in Ref.~\cite{evslin2019}.  Here the elementary
P\"oschl--Teller modes are replaced by the Heun-based spectral data
developed in Secs.~\ref{sec:heun} and \ref{sec:bound-states}.

\subsection{Diagonal quadratic Hamiltonian}
\label{subsec:diagonal-H}

We can now return to the quadratic Hamiltonian.  Using the definition
of $U_\epsilon$ in Eq.~\eqref{eq:U-definition}, it may be written as
\begin{equation}
H_K
=
H_0
-
\frac{1}{2}\int\dd x\,U_\epsilon(x):\phi^2(x):,
\label{eq:HK-free-plus-U}
\end{equation}
where $H_0$ is the spatial integral of
Eq.~\eqref{eq:H0-density}. When the transformations above are inserted into $H_0$, diagonal
oscillator terms appear together with terms that mix creation and
annihilation operators.  The latter cancel against the corresponding
terms generated by the localized potential.  This cancellation follows
from the fluctuation eigenvalue equation.
For a continuum mode we use
\begin{equation}
\int\dd x\,e^{-ipx}U_\epsilon(x)g_k(x)
=
\left(\omega_p^2-\omega_k^2\right)\widetilde g_k(-p),
\label{eq:Ug-continuum-identity}
\end{equation}
while for a bound state the analogous relation is
\begin{equation}
\int\dd x\,e^{-ipx}U_\epsilon(x)g_B(x)
=
\left(\omega_p^2-\omega_B^2\right)\widetilde g_B(-p).
\label{eq:Ug-bound-identity}
\end{equation}
The remaining frequency factors are simplified by
\begin{equation}
\frac{(\omega_p-\omega)^2}{\omega}
+
\frac{\omega_p^2-\omega^2}{\omega_p}
-
\frac{\omega_p^2-\omega^2}{\omega}
=
-\frac{(\omega_p-\omega)^2}{\omega_p},
\label{eq:frequency-identity}
\end{equation}
where $\omega$ may denote either a continuum or a nonzero bound-state
frequency.
The continuum contribution to the one-loop scalar term is therefore
\begin{equation}
Q_C
=
-\frac{1}{4}
\int\frac{\dd k}{2\pi}
\int\frac{\dd p}{2\pi}
\frac{(\omega_p-\omega_k)^2}{\omega_p}
\frac{|\widetilde g_k(p)|^2}{C_k^2}.
\label{eq:QC-general}
\end{equation}
Similarly, each nonzero bound state contributes
\begin{equation}
Q_B
=
-\frac{1}{4}
\int\frac{\dd p}{2\pi}
\frac{(\omega_p-\omega_B)^2}{\omega_p}
\frac{|\widetilde g_B(p)|^2}{C_B^2}.
\label{eq:QB-general}
\end{equation}
For the translational mode the oscillator description is replaced by
the collective coordinate.  Its scalar contribution is obtained from
the $\omega_B\to0$ limit of Eq.~\eqref{eq:QB-general},
\begin{equation}
Q_0
=
-\frac{1}{4}
\int\frac{\dd p}{2\pi}\,
\omega_p
\frac{|\widetilde g_0(p)|^2}{C_0^2}.
\label{eq:Q0-general}
\end{equation}

Combining the three sectors, we arrive at the diagonal quadratic
Hamiltonian
\begin{equation}
H_K
=
\int\frac{\dd k}{2\pi}\,\omega_k b_k^{\dagger}b_k
+
\sum_B\omega_B b_B^{\dagger}b_B
+
\frac{\pi_0^2}{2}
+
Q(\epsilon),
\label{eq:diagonal-HK}
\end{equation}
where
\begin{equation}
Q(\epsilon)
=
Q_C(\epsilon)
+
Q_0(\epsilon)
+
\sum_B Q_B(\epsilon).
\label{eq:Q-total-general}
\end{equation}
The rest energy is obtained by placing all nonzero-frequency
oscillators in their vacuum and taking vanishing collective momentum.
The latter is understood as a generalized state approached by localized
wave packets.  To one-loop order,
\begin{equation}
M_K
=
M_{\mathrm{cl}}
+
Q(\epsilon)
+
\mathcal{O}(\lambda/m).
\label{eq:kink-energy-general}
\end{equation}
It is useful to note that $U_\epsilon(x)$ no longer appears explicitly
in Eqs.~\eqref{eq:QC-general} and \eqref{eq:QB-general}.  Its effect is
contained entirely in the spectral data: the eigenfrequencies,
eigenfunctions, Fourier transforms, and normalization constants.  The
same structural simplification appears in the operator approach of
Ref.~\cite{evslin2019}.

\subsection{Renormalized one-loop correction and numerical evaluation}
\label{subsec:one-loop-evaluation}
\subsubsection{Trace representation and renormalization}
\label{subsec:epsilon-dependence}

The expressions above show separately how the continuum, bound states,
and zero mode enter the quantum correction.  For numerical work,
however, it is more convenient to combine all of them into a single
operator trace. We write
\begin{equation}
Q(\eps)=mF(\eps),
\end{equation}
where $F(\epsilon)$ is dimensionless.  Along with the kink operator
$\mathcal K_\epsilon$, we introduce the vacuum operator and the
corresponding positive frequency operators,
\begin{equation}
\mathcal K_0\equiv-\partial_X^2+1,
\qquad
\Omega_\eps\equiv\sqrt{\mathcal K_\eps},
\qquad
\Omega_0\equiv\sqrt{\mathcal K_0}.
\end{equation}
Using the completeness relation, the full mode sum becomes
\begin{equation}
\displaystyle
F(\eps)
=
\frac12\Tr(\Omega_\eps-\Omega_0)
+
\frac14\Tr(u_\eps\Omega_0^{-1}).
\label{eq:trace}
\end{equation}
The two terms in Eq.~\eqref{eq:trace} must be regulated together; they
are not separately finite traces.  The second term is precisely the
subtraction associated with vacuum normal ordering, or equivalently the
no-tadpole prescription used here.  We do not introduce an additional
finite on-shell counterterm.

This choice matters when numerical values from different calculations
are compared.  A finite change of renormalization scheme can shift the
one-loop mass, so both the vacuum mass convention and the subtraction
prescription must be held fixed
\cite{graham1998,bordag,graham2025}.  Throughout this paper,
``agreement'' therefore means agreement within the same
vacuum-normal-ordering prescription. The ultraviolet cancellation can also be seen directly from
Eq.~\eqref{eq:trace}.  At large free momentum, the first-order change
of the square-root operator contributes
$-u_\eps/(2\Omega_0)$ inside the first trace.  This is cancelled by the
normal-ordering term, leaving an integrable ultraviolet remainder in
one spatial dimension.  Appendix~\ref{app:trace} gives the regulated
derivation and also explains why the zero mode has to remain in the
completeness relation even though its direct contribution to
$\Tr\Omega_\eps$ vanishes.

\subsubsection{Independent imaginary-frequency determinant}
\label{sec:det}

The trace formula will provide our main numerical evaluation.  It is
nevertheless useful to have a second calculation that does not rely on
diagonalizing the same discretized operator.  For this purpose we use
an imaginary-frequency functional determinant. We introduce a dimensionless Euclidean frequency $\zeta>0$ and define
\begin{equation}
s\equiv\sqrt{1+\zeta^2}.
\end{equation}
To distinguish the determinant solution from the shifted classical kink
$f(x)$, we denote the decaying Jost solution by $J_s(X)$ and define it
through
\begin{equation}
J_s''(X)=[s^2-u_\eps(X)]J_s(X),
\qquad
J_s(X)\sim e^{-sX}
\quad(X\to+\infty).
\end{equation}
Since the fluctuation potential is even, the full-line determinant may
be constructed from the two half-line parity problems.  With the
normalization above,
\begin{equation}
D_\eps(\zeta)
=
\frac{J_s(0)[-J_s'(0)]}{s}.
\end{equation}
For the vacuum this quantity is equal to one.  The factors
$J_s(0)$ and $-J_s'(0)$ correspond to the usual Neumann and Dirichlet
half-line determinants in the Sturm--Liouville construction
\cite{kirsten,dunne}.

Using the integral representation of the square root in
Eq.~\eqref{eq:trace}, the same one-loop correction can be written as
\begin{equation}
F(\eps)
=
\frac1{2\pi}
\int_0^\infty\dd\zeta
\left[
\log D_\eps(\zeta)
+
\frac{I_\eps}{2\sqrt{1+\zeta^2}}
\right],
\qquad
I_\eps
=
\int_{\mathbb R}u_\eps(X)\,\dd X.
\label{eq:det}
\end{equation}
The second term removes the leading large-$\zeta$ Born contribution
and implements the same subtraction scheme as the trace formula.

The translational zero mode is also present in this representation.
Near the origin,
\[
D_\eps(\zeta)\propto\zeta^2.
\]
Although $\log D_\eps(\zeta)$ is singular as $\zeta\to0$, the
singularity is integrable and should not be removed by treating the
zero mode as a missing state. The determinant representation therefore gives a genuinely independent
numerical check.  It works directly with the differential equation for
$J_s$ and does not require Fourier transforms of the Heun modes.
Agreement between Eqs.~\eqref{eq:trace} and \eqref{eq:det} tests the
one-loop correction through two rather different numerical
constructions.

\subsubsection{Numerical implementation and convergence}

We first test the regulated trace numerically.  For
Eq.~\eqref{eq:trace}, we use a centered second-difference
discretization on the interval $[-L,L]$ with Dirichlet boundary
conditions.  The kink and vacuum operators are placed on exactly the
same grid, including the same lattice dispersion in the subtraction
term.  Our reference choice is
\[
L=24,
\qquad
h=0.08,\ 0.04,\ 0.02,\ 0.01.
\]
Because the continuum operator has an exact translational zero mode,
the lowest lattice eigenvalue should approach zero as the grid is
refined.  At finite $h$, however, the second-difference approximation
produces a small spurious negative value.  We identify this eigenvalue
with the analytic zero mode and set it to zero only before taking the
square root.  No negative excited eigenvalue is removed.

For example, at $\epsilon=1$ the uncorrected lowest lattice
eigenvalues are
\[
-1.9112\times10^{-5},\quad
-4.7776\times10^{-6},\quad
-1.1944\times10^{-6},\quad
-2.9859\times10^{-7}
\]
for $h=0.08,0.04,0.02,0.01$, respectively.  Their approximately
$h^2$ scaling confirms that they are discretization errors in the exact
zero mode rather than physical negative modes.

To extract the continuum value, we fit the lattice results to
\begin{equation}
F_h
=
F+c_1h^2\log h+c_2h^2.
\end{equation}
The three finest grids are used for the reference extrapolation, while
a four-grid fit provides a check on the sensitivity to the coarsest
point.  The logarithmic term is included to accommodate the ultraviolet
behavior of the second-difference regulator; we regard this as an
empirical extrapolation ansatz rather than a rigorous error expansion.

For $\eps\geq0.05$, the three-grid and four-grid fits differ by less
than $3\times10^{-8}$ in the sampled cases.  The softer spectra at
$\eps=0.01$ and $0.02$ require an additional $h=0.005$ grid, and we
quote those values more conservatively.  Increasing the box size from
$L=24$ to $L=32$ at $h=0.02$ changes $F_h$ by less than
$3\times10^{-10}$ in checks at $\eps=0.01,1,100$.

\begin{table}[H]
\centering
\caption{Convergence of the combined trace.  The exact translational
zero mode is imposed at each grid spacing.  The determinant values do
not depend on the spatial lattice.}
\label{tab:conv}
\begin{tabular}{@{}lrr@{}}
\toprule
Method & $F(1)$ & $F(\infty)$\\
\midrule
$h=0.08$ & $-0.389836962$ & $-0.333425536$\\
$h=0.04$ & $-0.389539677$ & $-0.333218435$\\
$h=0.02$ & $-0.389448710$ & $-0.333154197$\\
$h=0.01$ & $-0.389421821$ & $-0.333135030$\\
Three-grid extrapolation & $-0.389411014$ & $-0.333127261$\\
Determinant integral & $-0.389411020$ & $-0.333127262$\\
Exact P\"oschl--Teller value & --- & $-0.333127262$\\
\bottomrule
\end{tabular}
\end{table}
The convergence of the common-grid trace, together with the independent
determinant calculation and the exact large-deformation benchmark, is
summarized in Table~\ref{tab:conv}.  The extrapolated trace and
determinant values agree closely at $\epsilon=1$, while the
$\epsilon\to\infty$ result approaches the exact P\"oschl--Teller value. The large-deformation benchmark shown in Table~\ref{tab:conv} is
\begin{equation}
F_{\rm PT}
=
\frac{1}{4\sqrt3}
-
\frac{3}{2\pi}
=
-0.333127261978\ldots .
\label{eq:PT}
\end{equation}
This is the standard $\phi^4$ one-loop result in units of the vacuum
mass~\cite{dhn,evslin2019}.  Agreement with the P\"oschl--Teller limit
alone would not test the finite-deformation calculation, so we also
compare the trace and determinant representations away from this limit.
At $\epsilon=1$, the two methods give
\[
F(1)=-0.389411014
\qquad\text{and}\qquad
F(1)=-0.389411020,
\]
respectively. We therefore quote
\[
F(1)=-0.3894110
\]
with a numerical uncertainty of order $10^{-7}$ based on these
comparisons.  This should be understood as a numerical estimate rather
than a rigorous error bound.

The determinant check is not restricted to \(\epsilon=1\).  Table~\ref{tab:det-cross}
compares representative finite-deformation values obtained from the
extrapolated common-grid trace with a direct integration of
Eq.~\eqref{eq:det}.  The determinant calculation uses \(L=36\), splits
the \(\zeta\)-integral at \(0.1,1,10\), integrates to \(\zeta=60\),
and adds the leading asymptotic tail described in
Appendix~\ref{app:numerics}.  The agreement over both the long-cavity
and near-\(\phi^4\) regimes provides a stronger test than a single
benchmark point.

\begin{table}[H]
\centering
\caption{Independent finite-deformation check of the renormalized
one-loop shift.  The trace entries are shown to the precision used in the deformation
scan; the determinant column is obtained directly
from Eq.~\eqref{eq:det}.}
\label{tab:det-cross}
\begin{tabular}{@{}ccc@{}}
\toprule
$\epsilon$ & Extrapolated trace & Determinant \\
\midrule
$0.10$ & $-0.727170$ & $-0.72717035$ \\
$0.20$ & $-0.636688$ & $-0.63668847$ \\
$0.50$ & $-0.484589$ & $-0.48458895$ \\
$1.00$ & $-0.389411014$ & $-0.389411020$ \\
$2.00$ & $-0.348080$ & $-0.34807985$ \\
\bottomrule
\end{tabular}
\end{table}

\subsubsection{Dependence on the deformation parameter}

\begin{figure}[H]
\centering
\includegraphics[width=0.92\textwidth]{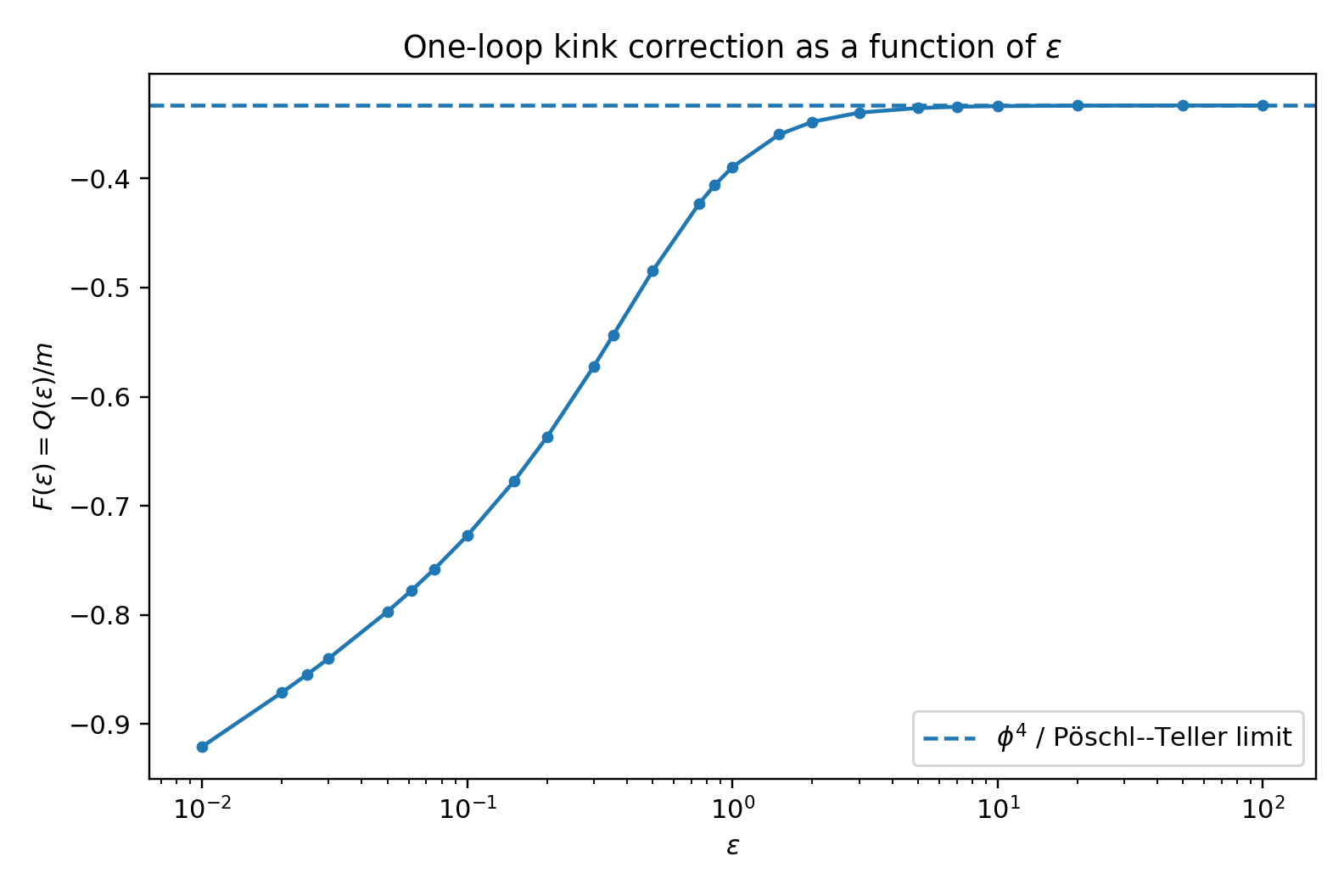}
\caption{Dimensionless one-loop correction
$F(\epsilon)=Q(\epsilon)/m$ as a function of the deformation parameter.
The solid curve connects the extrapolated numerical values and is
included as a guide to the eye.  The dashed horizontal line shows the
exact $\phi^4$/P\"oschl--Teller limit in Eq.~\eqref{eq:PT}.}
\label{fig:energy}
\end{figure}

Having established the numerical convergence, we now examine how the one-loop correction changes with \(\epsilon\). The resulting dependence is shown in Fig.~\ref{fig:energy}. Over the sampled range, the correction remains negative and becomes progressively less negative as \(\epsilon\) increases, approaching the exact \(\phi^4\)/Pöschl--Teller limit for large deformation. For reference, at \(\epsilon=1\) we obtain \(F(1)=-0.389411\).

The shallow even state discussed in Sec.~\ref{sec:shallow} approaches
the continuum threshold as $\epsilon\to\infty$, but it should not be
removed from the spectrum at any finite deformation.  Its contribution
is already part of the complete regulated trace.  The same point
applies to the finite threshold crossings listed in
Table~\ref{tab:thresholds}: when a mode crosses the threshold, its
contribution moves between the continuum and discrete sectors rather
than disappearing from the spectrum. This continuity is particularly transparent in the common-grid
calculation.  For fixed $L$ and $h$, the finite matrix
$\mathcal K_\epsilon$ varies smoothly with $\epsilon$, and therefore so
does the combined trace $F_h(\epsilon)$, even when an eigenvalue passes
through the finite-volume representation of the continuum threshold.
The division into ``bound'' and ``continuum'' pieces is thus
representation dependent at a crossing, whereas the regulated total is
not.  We do not, however, use this observation to claim differentiability
of the infinite-volume result at every threshold.

A sharper comparison can be made with the generalized heat-kernel
calculation of Ref.~\cite{alonso2011}.  That work uses
\begin{equation}
U_a(y)=\frac12(y^2+a^2)(1-y^2)^2
\end{equation}
and explicitly removes the one-loop ultraviolet divergence by normal
ordering.  Thus, after the vacuum-mass normalization is matched, the
renormalization prescription is the same no-tadpole prescription used
in Eq.~\eqref{eq:trace}; the difference between the two numerical
answers cannot be assigned to an arbitrary finite counterterm.

The precise conversion is simple.  In the notation of
Ref.~\cite{alonso2011} the vacuum fluctuation mass is
\(v_a=2\sqrt{1+a^2}\).  Setting \(a=\epsilon\) and rescaling their
coordinate by \(X=v_a x\) gives our unit-threshold operator.  Therefore
an energy shift quoted there in units of their scale \(m_d\) converts
to our convention according to
\begin{equation}
F_{\rm HK}(\epsilon)
=
\frac{1}{2\sqrt{1+\epsilon^2}}
\frac{\Delta E_{\rm HK}}{\hbar m_d}.
\label{eq:HK-conversion}
\end{equation}
Their Table~4 was evaluated with a finite truncation at eleven Seeley
coefficients.  Table~\ref{tab:heat-kernel-comparison} compares those
converted values with the present trace/determinant result.

\begin{table}[H]
\centering
\caption{Comparison with the finite heat-kernel truncation of
Ref.~\cite{alonso2011}.  The values in the second column are obtained by converting the $N = 11$ heat-kernel results $\Delta E/(\hbar m_d)$ reported in Table 4 of Ref. ~\cite{alonso2011} using Eq.~\eqref{eq:HK-conversion}.  The relative
difference is measured with respect to the present result.}
\label{tab:heat-kernel-comparison}
\begin{tabular}{@{}cccc@{}}
\toprule
$\epsilon$ & $F_{\rm HK}^{(N=11)}$ & Present $F(\epsilon)$ & Relative difference \\
\midrule
$0.10$ & $-0.671192$ & $-0.727170$ & $7.70\%$ \\
$0.20$ & $-0.607587$ & $-0.636688$ & $4.57\%$ \\
$0.30$ & $-0.553954$ & $-0.572048$ & $3.16\%$ \\
$0.50$ & $-0.478036$ & $-0.484589$ & $1.35\%$ \\
$1.00$ & $-0.389181$ & $-0.389411$ & $0.059\%$ \\
\bottomrule
\end{tabular}
\end{table}

The pattern is systematic: the finite heat-kernel estimate is very
close near \(\epsilon=1\), while the difference grows as the kink
separates into two interfaces.  Ref.~\cite{alonso2011} explicitly
emphasizes that its heat-kernel formula is an asymptotic series whose
accuracy depends on the truncation order.  The long-cavity regime adds
a second large geometric scale, so a fixed local asymptotic truncation
need not be uniform as \(\epsilon\to0^+\).  By contrast, the present
common-grid trace and the independent determinant calculation agree at
\(\epsilon=0.1\), $F(0.1)=-0.72717035$,
and similarly at the representative points in
Table~\ref{tab:det-cross}.  We therefore interpret the difference in
Table~\ref{tab:heat-kernel-comparison} as a truncation issue in the
previous approximate heat-kernel evaluation, rather than a
renormalization-scheme ambiguity.  Establishing the detailed
large-order behavior of the Seeley expansion in the two-interface
limit would be an interesting separate problem.

\subsubsection{Asymptotic deformation regimes of the one-loop mass}
\label{subsubsec:asymptotic-energy}

\paragraph{Large-$\epsilon$ behavior.}
\label{subsec:large-energy}

The numerical trend in Fig.~\ref{fig:energy} suggests a smooth approach
to the $\phi^4$ result.  The large-$\epsilon$ operator expansion derived
in Sec.~\ref{sec:shallow} allows us to make this behavior more
quantitative. Writing $d=\epsilon^{-2}$, we have
\begin{equation}
\mathcal K_\epsilon
=
\mathcal K_\infty+dW_1+O(d^2),
\qquad
u_\epsilon
=
u_\infty-dW_1+O(d^2).
\end{equation}
If we formally differentiate the already-subtracted trace, the
first-order change is
\begin{equation}
\delta F^{(1)}
=
\frac14\,\Tr{}'\!\left[
W_1
\left(
\mathcal K_\infty^{-1/2}
-
\mathcal K_0^{-1/2}
\right)
\right],
\label{eq:Cinf-trace}
\end{equation}
where the prime indicates that the translational zero mode is treated
by continuity, since its eigenvalue remains exactly zero throughout the
deformation. This gives the asymptotic form
\begin{equation}
F(\epsilon)
=
F_{\rm PT}
-
\frac{C_\infty}{\epsilon^2}
+
O(\epsilon^{-4}),
\qquad
C_\infty=-\delta F^{(1)}.
\label{eq:large-energy-asymptotic}
\end{equation}
The numerical scan provides a direct check.  We find
\begin{equation}
-\epsilon^2
\bigl[F(\epsilon)-F_{\rm PT}\bigr]
=
0.057900,\quad
0.057776,\quad
0.057738,\quad
0.057720
\end{equation}
for $\epsilon=10,20,50,100$, respectively.  A two-term fit in
$\epsilon^{-2}$ and $\epsilon^{-4}$ gives
\begin{equation}
C_\infty=0.0577352
\qquad
\text{(numerical fit)}.
\label{eq:Cinf-numerical}
\end{equation}
We do not assign an exact closed form to this number without evaluating
Eq.~\eqref{eq:Cinf-trace} analytically.  One point is nevertheless
already clear: the leading change in the full one-loop energy is
$O(\epsilon^{-2})$, whereas the frequency gap of the shallow even mode
found in Eq.~\eqref{eq:shallow} is only $O(\epsilon^{-4})$.  The
approach of the kink mass to its $\phi^4$ value is therefore controlled
by the deformation of the complete fluctuation spectrum, not by the
near-threshold state alone.

\paragraph{Small-$\epsilon$ behavior.}
\label{sec:small}

As shown in Sec.~\ref{subsec:small-spectrum}, in the limit $\epsilon\to0^+$ the two
interfaces separate according to
\begin{equation}
\ell_X
=
4\log\frac{2}{\epsilon}
+
O(1),
\end{equation}
while the dimensionless fluctuation operator approaches
$-\partial_X^2+1/4$ throughout the increasingly long central region.
We now consider the consequence of this geometry for the one-loop
energy.

For a uniform region with dimensionless mass squared $a$, normal
ordered with respect to the asymptotic mass squared one, the
renormalized energy density is
\begin{equation}
\mathcal E(a)
=
\frac12
\int_{\mathbb R}\frac{\dd p}{2\pi}
\left[
\sqrt{p^2+a}
-
\sqrt{p^2+1}
-
\frac{a-1}{2\sqrt{p^2+1}}
\right].
\label{eq:density}
\end{equation}
The subtraction is exactly the same one that appears in
Eq.~\eqref{eq:trace}. Differentiating with respect to $a$ gives
\begin{equation}
\mathcal E'(a)
=
-\frac{\log a}{8\pi},
\qquad
\mathcal E(1)=0,
\end{equation}
and hence
\begin{equation}
\mathcal E(a)
=
\frac{a-1-a\log a}{8\pi},
\qquad
\mathcal E(1/4)
=
\frac{\log4-3}{32\pi}
<0.
\end{equation}
The central region has length
$\ell_X=4\log(2/\epsilon)+O(1)$, so multiplying this energy density by
the cavity length gives the leading prediction
\begin{equation}
F(\eps)
=
-\frac{3-\log4}{8\pi}
\log\frac2\eps
+
O(1),
\qquad
\eps\to0^+.
\label{eq:small}
\end{equation}
Equation~\eqref{eq:small} should be interpreted as a leading bulk
asymptotic rather than as a complete theorem for the remainder.  The
coefficient of $\log(2/\epsilon)$ follows directly from the
renormalized vacuum-energy density of the long central region.  The
statement that the remaining contribution is $O(1)$ additionally
assumes that the two transition regions approach fixed interface
profiles with finite renormalized interface energies.

Collecting the classical and one-loop contributions, the
vacuum-subtracted kink rest mass is
\begin{equation}
M_K(\epsilon)
=
M_{\mathrm{cl}}(\epsilon)
+
mF(\epsilon)
+
O(\lambda/m),
\qquad
\epsilon>0,
\end{equation}
with $M_{\mathrm{cl}}$ given in Eq.~\eqref{eq:classical}.  All numerical
mass shifts quoted above refer to the fixed vacuum-normal-ordering
prescription used throughout this section.

% ============================================================
% ============================================================
\section{Conclusions}
\label{sec:conclusion}

We have developed a spectral description of the Christ--Lee kink that
connects the analytic structure of the fluctuation equation with the
deformation dependence of its one-loop quantum mass. The fluctuation
problem reduces exactly to general-Heun form, while the global spectrum
is obtained independently by direct shooting method. In particular, the
terminating odd Heun solution yields the exact first nontrivial
threshold,$\epsilon_1=\sqrt{\sqrt{3}-1}$,
providing an analytic benchmark for the numerical spectrum.

The two deformation limits exhibit distinct spectral mechanisms. At
small $\epsilon$, the kink separates into two $\phi^6$-like interfaces
and develops a long central region that supports a cavity-like tower of
bound states together with a soft relative-translation mode. At large
$\epsilon$, the spectrum approaches the $\phi^4$ P\"oschl--Teller
problem, with a shallow even state emerging from the threshold
resonance of the limiting theory. The renormalized one-loop energy was evaluated using both a
common-regulator spectral trace and an independent imaginary-frequency
functional determinant. Their agreement at finite deformation and the
recovery of the exact $\phi^4$ limit provide nontrivial checks of the
calculation. The comparison with the finite heat-kernel approximation
shows excellent agreement near the $\phi^4$ regime, while deviations
increase as the two-interface separation becomes large.

The main result is that the quantum-mass dependence on $\epsilon$ is
controlled by the reorganization of the complete fluctuation spectrum.
At small deformation, the extended central region produces the leading
logarithmic enhancement of the negative one-loop correction, whereas at
large deformation the quantum mass approaches the $\phi^4$ value
smoothly despite the presence of a near-threshold state. The
semiclassical expansion consequently becomes nonuniform in the
widely-separated-interface limit. Future work should include a systematic high-order analysis of the
heat-kernel expansion in the small-$\epsilon$ regime and an extension
of the semiclassical treatment to higher-loop effects, particularly in
the presence of the soft relative-translation mode.
\newpage
\appendix
\section{Direct verification of the parity Heun modes}
\label{app:verification}

This appendix records a compact direct check of Eqs.~\eqref{eq:even-continuum} and \eqref{eq:odd-continuum}.  With the variables in Eq.~\eqref{eq:heun-variables}, define
\begin{equation}
R(z)
=
\frac{3}{4}
\frac{(1+\epsilon^2)(1+2\epsilon^2)-2\epsilon^2(\epsilon^2+3)z}
{(1+\epsilon^2-\epsilon^2z)^2}.
\label{eq:appendix-R}
\end{equation}
The original fluctuation operator in Eq.~\eqref{eq:fluctuation-equation} is then
\begin{equation}
\mathcal{L}[G]
=
m^2\left[
z(z-1)G_{zz}
+
\frac{2z-1}{2}G_z
+
\left(\nu^2+R(z)\right)G
\right].
\label{eq:appendix-L}
\end{equation}
For the even mode, set
\begin{equation}
G(z)=P(z)H_e(z),
\qquad
P(z)=\left(\frac{a}{a-z}\right)^{3/2}.
\label{eq:appendix-even-factor}
\end{equation}
The logarithmic derivatives of $P$ are
\begin{equation}
\frac{P'}{P}=-\frac{3}{2(z-a)},
\qquad
\frac{P''}{P}=\frac{15}{4(z-a)^2}.
\label{eq:appendix-P-derivatives}
\end{equation}
Substitution of Eqs.~\eqref{eq:appendix-even-factor} and \eqref{eq:appendix-P-derivatives} into Eq.~\eqref{eq:appendix-L} yields
\begin{equation}
\begin{aligned}
\frac{\mathcal L[P H_e]}{m^2Pz(z-1)}
={}&H_e^{\prime\prime}
+\left[\frac{1}{2z}+\frac1{2(z-1)}-\frac3{z-a}\right]H_e^\prime\\
&+\frac{(\nu^2+9/4)z-a(\nu^2+3/4)}{z(z-1)(z-a)}H_e.
\end{aligned}
\label{eq:appendix-even-verification}
\end{equation}
The right-hand side of Eq.~\eqref{eq:appendix-even-verification} is precisely the Heun equation in Eq.~\eqref{eq:heun-reduced}; hence $\mathcal{L}[E_k]=0$.

For the odd mode on $x>0$, write
\begin{equation}
G(z)=Q(z)H_o(z),
\qquad
Q(z)=P(z)(-z)^{1/2}.
\label{eq:appendix-odd-factor}
\end{equation}
Using Eq.~\eqref{eq:appendix-odd-factor} in Eq.~\eqref{eq:appendix-L} gives
\begin{equation}
\begin{aligned}
\frac{\mathcal L[Q H_o]}{m^2Qz(z-1)}
={}&H_o^{\prime\prime}
+\left[\frac{3}{2z}+\frac1{2(z-1)}-\frac3{z-a}\right]H_o^\prime\\
&+\frac{(\nu^2+1)z-a(\nu^2+1)+3/2}{z(z-1)(z-a)}H_o.
\end{aligned}
\label{eq:appendix-odd-verification}
\end{equation}
Equation~\eqref{eq:appendix-odd-verification} is the canonical Heun equation with
\begin{equation}
\alpha_o=-1+i\nu,
\qquad
\beta_o=-1-i\nu,
\qquad
\gamma_o=\frac{3}{2},
\qquad
\delta_o=\frac{1}{2},
\qquad
q_o=a(\nu^2+1)-\frac{3}{2}.
\label{eq:appendix-odd-parameters}
\end{equation}
Thus $\mathcal{L}[O_k]=0$.  The local behavior $z=-m^2x^2/4+\mathcal{O}(x^4)$ also reproduces the origin conditions in Eqs.~\eqref{eq:even-origin} and \eqref{eq:odd-origin}, completing the direct verification. The extension to $x<0$ is fixed by odd parity, rather than by a positive square root of $-z$.

% ============================================================
\section{Reduction of the mode sum to a trace}\label{app:trace}
Work first in a common finite-dimensional regulator. Let $\mathcal K_\eps|n\rangle=\omega_n^2|n\rangle$ and $\mathcal K_0|p\rangle=\omega_p^2|p\rangle$, now with dimensionless frequencies. Include the zero eigenvector in the index $n$. The continuum, bound-state and zero-mode contributions combine to
\begin{equation}
F=-\frac14\sum_{n,p}\left(\omega_p-2\omega_n+
\frac{\omega_n^2}{\omega_p}\right)|\langle p|n\rangle|^2.
\end{equation}
Completeness gives
\begin{equation}
F=-\frac14\Tr\Omega_0+\frac12\Tr\Omega_\eps
-\frac14\Tr(\mathcal K_\eps\Omega_0^{-1}).
\end{equation}
Since $\mathcal K_\eps=\mathcal K_0-u_\eps$ and $\mathcal K_0\Omega_0^{-1}=\Omega_0$, this is precisely~\eqref{eq:trace}. No commutation between $u_\eps$ and $\Omega_0$ is assumed. At finite dimension all sums are ordinary sums; the regulator is removed only after the three terms have been combined. A factorization-preserving discretization can keep the zero mode exact. In the second-difference implementation we instead reset its converging spurious eigenvalue to zero, as described in the main text, before taking the continuum limit.

\section{Numerical implementation and reproducibility}\label{app:numerics}
For $N$ interior Dirichlet points and $h=2L/(N+1)$, the matrices are
\begin{equation}
(\mathcal K_\eps)_{ij}=\left(\frac2{h^2}+1-u_\eps(X_i)\right)\delta_{ij}
-\frac{\delta_{i,j+1}+\delta_{i,j-1}}{h^2}.
\end{equation}
The free eigenvalues and eigenvectors are known exactly:
\begin{equation}
\lambda_j^{(0)}=1+\frac4{h^2}\sin^2\frac{j\pi}{2(N+1)},\qquad
v_j(i)=\sqrt{\frac2{N+1}}\sin\frac{ij\pi}{N+1}.
\end{equation}
Thus the counterterm uses
\begin{equation}
(\Omega_0^{-1})_{ii}=\frac2{N+1}\sum_{j=1}^N
\frac{\sin^2[ij\pi/(N+1)]}{\sqrt{\lambda_j^{(0)}}}.
\end{equation}
The discrete trace is an ordinary matrix trace with this normalization; no extra factor of $h$ is inserted. A fast Fourier transform evaluates the cosine part of $2\sin^2\theta=1-\cos2\theta$. The kink eigenvalues are obtained from a real symmetric tridiagonal solver. The accompanying Python code records raw grid values, fits, bound-state roots and thresholds.

For the determinant, define $r=J_s'/J_s+s$ and $b=\log(J_se^{sX})$. They obey
\begin{equation}
r'=2sr-r^2-u_\eps,\qquad b'=r,
\qquad r(L)=b(L)=0,
\end{equation}
so that $\log D=2b(0)+\log[1-r(0)/s]$. Inward integration suppresses the unwanted exponential branch. The check uses $L=36$, relative tolerance $2\times10^{-11}$ and absolute tolerance $2\times10^{-13}$. The integral is split at $\zeta=0.1,1,10$ and terminated at 60. Near zero we use the integrable form $\log D=2\log\zeta+O(1)$ below $\zeta_0=3\times10^{-4}$. The leading tail beyond $T$ is
\begin{equation}
-\frac{1}{32\pi T^2}\int_{\mathbb R}u_\eps^2\,\dd X,
\end{equation}
with higher terms suppressed by additional inverse powers of $T$. The exact P\"oschl--Teller value tests the complete procedure, including both endpoints. These checks support the quoted precision but do not supply a rigorous a priori error theorem.

\section*{Acknowledgments}
This work was supported by the BRIN Postdoctoral Research Program 2026. B.E.G. also acknowledges financial support from the PPMI KK FMIPA ITB No. 062/IT1.C02/SK-DA/2026.

\section*{Data availability statement}
No external observational or experimental datasets were used. The numerical results reported in this article were generated directly from the equations specified in the manuscript.

\section*{Conflict of interest}
The author declares no conflict of interest.

\end{document}